\documentclass[a4paper,11pt]{article}
\pdfoutput=1 

\usepackage{jcappub} 

\usepackage[T1]{fontenc} 

\usepackage{relsize}
\usepackage[compat=1.0.0]{tikz-feynman}
\usepackage{comment}
\usepackage{subcaption}
\newcommand{\ki}{\mathbf{k}}
\newcommand{\qi}{\mathbf{q}}

\newcommand{\D}{\mathrm{d}}

\title{\boldmath Effects of scalar non-Gaussianity on induced scalar-tensor gravitational waves}

\author{Rapha{\"e}l Picard}
\author{and Matthew W. Davies}

\affiliation{Astronomy Unit, Queen Mary University of London, \\
Mile End Road, London, E1 4NS, UK}

\emailAdd{r.h.j.picard@qmul.ac.uk}
\emailAdd{m.w.davies@qmul.ac.uk}

\abstract{If primordial scalar or tensor perturbations are enhanced on short scales, it may lead to the production of observable gravitational wave signals. These waves may be sourced by scalar-scalar, scalar-tensor or tensor-tensor interactions. Typically, models of inflation capable of producing large peaks in the scalar primordial power spectrum also generate sizeable scalar non-Gaussianity. Previous studies have investigated the possible effects of this on the scalar-scalar induced gravitational wave spectrum by assuming a local expansion in terms of the parameters $F_{\textrm{NL}}$, $G_{\textrm{NL}}$ and so on. We extend this approach to the case of scalar-tensor induced gravitational waves, introducing a local expansion for scalar non-Gaussianity into the scalar-tensor sector equations. We compute the contribution to the gravitational wave spectrum from the resulting new term and analyse its distinguishing features.}

\begin{document}
\maketitle
\flushbottom
\section{Introduction} \label{sec:intro}
The first direct detection of gravitational waves (GWs) by the LIGO and Virgo collaborations in 2016~\cite{LIGOScientific:2016aoc} has opened up a brand new window through which to view universe. Since this detection many more GW events have been observed, and the future of GW cosmology looks very promising. Some of the most recent achievements in GW observation include the detections of mergers of compact objects~\cite{LIGOScientific:2016aoc,KAGRA:2021vkt} and the recent data collected from pulsar timing array (PTA) measurements which tantalisingly suggest the existence of a stochastic gravitational wave background (SGWB)~\cite{NANOGrav:2023gor,EPTA:2023sfo,EPTA:2023fyk,EPTA:2023xxk,Xu:2023wog,Reardon:2023gzh,Zic:2023gta,NANOGrav:2023hvm}. In addition, there are a number of proposals for space-based GW observatories that have been put forward that would enable us to probe GW signals at lower frequencies than ground based detectors. This includes LISA~\cite{Bartolo:2016ami}, DECIGO and BBO~\cite{Kawamura:2020pcg}. These experiments would offer a wealth of information concerning the physics of the early universe and could place constraints on inflationary scenarios which predict a SGWB. For a roadmap of future GW detectors and their sensitivity curves, see Refs.~\cite{Bailes:2021tot,Moore:2014lga}.

Of particular interest to this work, second-order gravitational waves (SOGWs) are GWs that appear at second order in cosmological perturbation theory and are sourced by first order scalar and tensor perturbations. SOGWs sourced only by scalar perturbations are termed scalar induced gravitational waves (SIGWs). The seeds for their study were originally sown by Tomita in 1967~\cite{Tomita:1967wkp} and Matarrese et. al. in 1992~\cite{Matarrese:1992rp}. Since then, SIGWs have received plenty of attention as the theory behind them was developed extensively in Refs.~\cite{ Matarrese:1992rp, Matarrese:1993zf, Matarrese:1997ay, Ananda:2006af, Baumann:2007zm}. Focus on SIGWs has intensified recently due to their links to primordial black holes (PBHs) --- detectable SIGW signals could indicate their existence. PBHs are formed when large density fluctuations generated by inflation re-enter the horizon during the radiation dominated era and collapse. This formation process typically requires an enhancement of the scalar primordial power spectrum of $\sim 10^7$~\cite{Motohashi:2017kbs} on short scales, relative to the amplitude $\mathcal{A}_s=2.1 \times 10^{-9}$ measured via the CMB~\cite{Planck:2018vyg}. An amplification this large would inevitably generate observable SIGWs as a counterpart signal. See Ref.~\cite{Domenech:2021ztg} for a review.

There has also been plenty of interest in the possible impact of primordial non-Gaussianity (PNG) on any potentially observable SIGW signal~\cite{Cai:2018dig,Unal:2018yaa,Yuan:2020iwf,Atal:2021jyo,Adshead:2021hnm,Ragavendra:2020sop,Zeng:2024ovg,Pi:2024jwt,Perna:2024ehx,Li:2023xtl}. PNG arises when the higher-order correlations of the curvature perturbation become non-zero, and can be seen as a measure of the interactions of the inflaton field beyond the free-field dynamics. This means that any detection of PNG would be particularly enlightening for cosmology, since it would tell us a wealth of information about the inflationary action beyond second order. At present, the constraints on non-Gaussianity are fairly weak; the most recent Planck data from CMB measurements gives the local non-Gaussianity as $f_{\textrm{NL}}^{\textrm{local}}=-0.9\pm5.1$~\cite{Planck:2019kim}. The impact of PNG on potentially observable GW signals is particularly pertinent since models of inflation capable of generating enhanced peaks, and therefore PBHs and SIGWs, typically involve curvature perturbations with non-Gaussian statistics, see e.g. Refs.~\cite{Cai:2021zsp,Davies:2021loj,Taoso:2021uvl,Rezazadeh:2021clf}. So far, works considering non-Gaussian scalar curvature perturbations have modelled the non-Gaussianity as a local expansion. It has been found that the introduction of scalar non-Gaussianity typically results in the smoothing of peaks and the appearance of `knees' at higher frequencies in the GW spectrum~\cite{Unal:2018yaa,Atal:2021jyo,Adshead:2021hnm,Pi:2024jwt,Perna:2024ehx,Li:2023xtl}.

While SOGWs induced by enhanced scalar perturbations have received the most attention, it is also possible for SOGWs to form from enhanced tensor perturbations through scalar-tensor or tensor-tensor interactions~\cite{Picard:2023sbz,Bari:2023rcw, Gong:2019mui,Chang:2022vlv,Yu:2023lmo}. Indeed, on small scales, the scalar-tensor channel of production becomes dominant. Although Planck data indicates that both the scalar amplitude ($\mathcal{A}_s=2.1 \times 10^{-9}$) and scalar-to-tensor ratio ($r<0.065$)~\cite{Planck:2018vyg}, and hence the tensor amplitude, are far too small on large scales to generate observable SOGWs, this need not be the case on short scales. Multifield models of
inflation, containing a spectator axion coupled
to an $SU(2)$ gauge field for example, can produce a peak in the tensor power spectrum at short scales~\cite{Barnaby:2011qe,Thorne:2017jft,Dimastrogiovanni:2016fuu}. It may be the case, therefore, that both scalar and tensor perturbations become enhanced on short scales, and that the contributions from scalar-tensor and tensor-tensor interactions to the SOGW spectrum must be taken into account. 

Since we would expect sizeable scalar non-Gaussianity to accompany an enhanced scalar power spectrum, the contribution to the SOGW spectrum from scalar-tensor interactions is likely to be affected by PNG. To our knowledge, this possibility has not yet been explored and so provides the motivation for this work. The paper is structured as follows. In Sec.~\ref{sec:SOGW} we recall how to compute the power spectrum of SOGWs splitting contributions into scalar-scalar and scalar-tensor interaction modes. In Sec.~\ref{NGsection} we locally expand the scalar curvature perturbation as a Gaussian and non-Gaussian part and introduce it into the formula for the scalar-tensor contribution. This yields a new non-Gaussian term in the scalar-tensor induced GW spectrum. In Sec.~\ref{resultsSec} we analyse the behaviour of the new term in the context of peaked primordial scalar and tensor power spectra in a radiation dominated universe. We compare the effect of this contribution to the non-Gaussian SIGW terms studied previously. We discuss our results and conclude in Sec.~\ref{sec:conclusions}.

\medskip
\textit{Conventions:} Conformal time is defined as $\mathrm{d}\eta \equiv \mathrm{d}t/a(t)$ and $f'\equiv {\mathrm{d}f}/{\mathrm{d}\eta}$. We use natural units and set the reduced Planck mass, $M_\text{Pl}\equiv(8\pi G_N)^{-1/2}$, to unity unless otherwise stated.

\section{Second order gravitational waves from scalar-scalar and scalar-tensor interactions}
\label{sec:SOGW}
We apply the conformal Newtonian gauge to a perturbed FLRW background to give the line element
\begin{equation}
\label{PerturbedFLRW}
    \D s^2 = a^2(\eta)\left [-(1+2\Phi^{(1)}  )\,\D\eta ^2+\left ((1-2\Psi^{(1)} )\delta _{ij}  +2\overline{h}^{(1)}_{ij}+\overline{h}^{(2)}_{ij} \right )\,\D x^i \D x^j \right]\, ,
\end{equation}
where $\eta$ and $a(\eta)$ are the conformal time and scale factor respectively. The functions $\Phi^{(n)}$ and $\Psi^{(n)}$ are the lapse and curvature perturbations to the metric at n-th order. The tensorial perturbations to the metric at n-th order are denoted by $\overline{h}_{ij}^{(n)}$ and are transverse-traceless (TT), i.e. $\partial^i\overline{h}_{ij}^{(n)}=\delta^{ij}\overline{h}_{ij}^{(n)}=0$. Note that first order vector perturbations are discarded since they are diluted away during inflation. Similarly, we do not consider second order scalars and vectors as they do not contribute to the second order tensor equation of motion. For simplicity, in what follows, the overline denoting TT quantities will be dropped and the order of scalar and tensor perturbations will only be displayed for quantities beyond first order.

After inflation, the content of our universe can be modelled as a perfect (barotropic) fluid with adiabatic initial conditions and a constant equation of state\footnote{This implies that the equation of state, $w$, and adiabatic sound speed, $c_s^2$, are equal. They are defined via
\begin{equation*}
    P=w\rho \, \, \text{and} \, \,  c_s^2=\frac{\partial P}{\partial \rho} \bigg |_{\delta S =0} \, .
\end{equation*}
where $P,\, \rho \, \, \text{and} \, \, S$ are the pressure, density and entropy repsectively.}. The stress-energy tensor, $T_{\mu \nu}$, can then be expressed as
\begin{equation}
\label{EMtensor}
    T_{\mu \nu} = (\rho + P ) u_{\mu} u_{\nu} + P g_{\mu \nu}\, ,
\end{equation}
where $u^{\mu}$ is a comoving $4$-velocity. Furthermore, we will ignore anisotropic stress at all orders as it leaves negligible imprint on the power spectrum of SIGWs \cite{Baumann:2007zm}.

\subsection{The SOGW equation and its solution}
The equation of motion for second-order gravitational waves is obtained by extracting the TT part of the spatial second-order Einstein equations. After some simplification using first-order equations of motion, the resulting equation is (see. Ref.~\cite{Picard:2023sbz} for details)
\begin{equation} 
\label{GW2eq}
    h_{ab}^{\prime \prime (2)} + 2\mathcal{H}h_{ab}^{\prime (2)} - \nabla ^2h_{ab}^{(2)}= \Lambda ^{ij}_{ab}(S^{ss} _{ij} + S^{st} _{ij})\, ,
\end{equation}
where $\mathcal{H}(\eta)\equiv a'(\eta)/a(\eta)$ is the conformal Hubble parameter, $\Lambda ^{ij}_{ab}$ is the transverse-traceless operator defined as 
\begin{equation}
    \label{TTprojector}
    \Lambda^{ij}_{ab}=\left(\delta^i_a-\frac{\partial^i\partial_a}{\nabla^2}\right)\left(\delta^j_b - \frac{\partial^j \partial_b}{\nabla^2}\right)-\frac{1}{2}\left(\delta_{ab}-\frac{\partial_a \partial_b}{\nabla^2}\right)\left(\delta^{ij}-\frac{\partial^i\partial^j}{\nabla^2}\right) \, ,
\end{equation}
 where the Laplacian in the denominator represents the inverse Laplacian defined according to $\nabla^{-2}(\nabla^2 X)=X$ and
\begin{subequations}
\label{sourceterms}
\begin{align}
\label{scalarsource}
S^{ss} _{ij} & = \frac{8}{3(1+w)}\left [(\partial _i\Psi +\frac{\partial _i \Psi ^{\prime}}{\mathcal{H}})(\partial _j \Psi +\frac{\partial _j \Psi ^{\prime}}{\mathcal{H}})\right ] +4 \partial _i \Psi \partial _j \Psi \,,
\\
\label{sclartensorsource}
\begin{split}
    S^{st}_{ij} & = 8\Psi\nabla^2h_{ij} + 8\partial _ch_{ij}\partial ^c \Psi + 4h_{ij} (   \mathcal{H}(1+3c_s^2) \Psi^{\prime} +(1-c_s^2)\nabla^2\Psi )\,,
\end{split}
\end{align}
\end{subequations}
are the source terms arising from the quadratic scalar-scalar and mixed scalar-tensor contributions respectively. We have neglected quadratic tensor-tensor perturbations in our analysis since they have a subdominant contribution to the SGWB in the context of primordial peaked sources~\cite{Picard:2023sbz}. 

To solve Eq.~\eqref{GW2eq} we transform to Fourier space. The scalar and tensor perturbations in real-space can be expressed as the Fourier integrals
\begin{subequations}
    \begin{align}
        \Psi(\mathbf{x},\eta)&= \int \frac{\D^3\mathbf{k}}{(2\pi)^{\frac{3}{2}}} \Psi(\mathbf{k},\eta)e ^{i\mathbf{k}\cdot \mathbf{x}}\,, \\
        h^{(n)}_{ab}(\mathbf{x},\eta) &=\int \frac{\D^3\mathbf{k}}{(2\pi)^{\frac{3}{2}}}  \left \{ h^{(n)}_{R}(\mathbf{k}, \eta) q_{ab}^{R}(\mathbf{k}) + h^{(n)}_{L}(\mathbf{k}, \eta) q_{ab}^{L}(\mathbf{k})  \right \}e ^{i\mathbf{k}\cdot \mathbf{x}} \,.
    \end{align}
\end{subequations}
Note that we work with circular polarisations, $\lambda=R/L$, and the associated polarisation tensors $q_{a b}^R(\mathbf{k})$ and $q_{a b}^L(\mathbf{k})$ are given in App.~\ref{sec:Polarisation}. Furthermore, perturbations can be split up in the following way
\begin{subequations}\label{defsplitfluctuations}
\begin{align}
\Psi(\eta , \mathbf{k})&= \left (\frac{3+3w}{5+3w}\right )T_{\Psi}(c_s\eta k)\mathcal{R}_{\mathbf{k}}\,, \\
h^{\lambda _1}(\eta , \mathbf{k}) &=T_{h}(\eta k)h^{\lambda _1}_{\mathbf{k}} \, ,
\end{align}
\end{subequations}
where $T_{\Psi}(c_s\eta k)$ and $T_{h}(\eta k)$ represent the transfer functions associated with the scalar and tensor perturbations respectively, and ${\mathcal{R}}_{\mathbf{k}}$ and $h^{\lambda _1}_{\mathbf{k}}$ represent the superhorizon values of the scalar and tensor perturbations, which are fixed at the end of inflation. The transfer functions thus encode the linear evolution of these perturbations from the point at which these scales re-enter the horizon after inflation. The transfer function for the tensor modes is independent of the polarisation $\lambda _1=R/L$, however the primordial values of $h_{\mathbf{k}}^R$ and $h_{\mathbf{k}}^L$ can differ.

As a result of transforming to Fourier space and splitting up the first order perturbations, Eq.~\eqref{GW2eq} now becomes
\begin{equation} 
\label{GW2fourier}
    h_{\lambda}^{(2)\prime \prime} (\eta , \mathbf{k}) + 2\mathcal{H}h_{\lambda}^{(2)\prime} (\eta , \mathbf{k}) +k^2h_{\lambda}^{(2)} (\eta , \mathbf{k}) =  4\left ( \mathcal{S}^{ss}_{\lambda} (\eta , \mathbf{k})+\mathcal{S}^{st}_{\lambda} (\eta , \mathbf{k})\right )\, .
\end{equation}
The form for the source terms, $\mathcal{S}^{ss}_{\lambda} (\eta , \mathbf{k})$ and $\mathcal{S}^{st}_{\lambda} (\eta , \mathbf{k})$, in Fourier space can be shown to be 
\begin{subequations}
\label{sourcetermsfourier}
\begin{align}
\label{scalarsourcefourier}
\mathcal{S}^{ss}_{\lambda} (\eta , \mathbf{k}) & = \int \frac{\D^3\mathbf{p}}{(2\pi)^{\frac{3}{2}}} Q^{ss}_{\lambda}(\mathbf{k},\mathbf{p})f^{ss}(c_s\eta p,c_s\eta |\mathbf{k}-\mathbf{p}|)\mathcal{R}_{\mathbf{p}} \mathcal{R}_{\mathbf{k}-\mathbf{p}} \,,
\\
\label{sclartensorsourcefourier}
\begin{split}
   \mathcal{S}^{st}_{\lambda} (\eta , \mathbf{k}) & =\sum_{\lambda _1 =R,L} \int \frac{\D^3\mathbf{p}}{(2\pi)^{\frac{3}{2}}} Q^{st}_{\lambda , \lambda _1}(\mathbf{k},\mathbf{p})f^{st}(\eta p,c_s\eta|\mathbf{k}-\mathbf{p}|)h_{\mathbf{p}}^{\lambda _1} \mathcal{R}_{\mathbf{k}-\mathbf{p}}\,,
\end{split}
\end{align}
\end{subequations}
where the functions $f^{ss}(c_s\eta p,c_s\eta |\mathbf{k}-\mathbf{p}|)$ and $f^{st}(\eta p,c_s\eta |\mathbf{k}-\mathbf{p}|)$ contain all the information about the subhorizon evolution of the perturbations and are given by
\begin{subequations}\label{transferfuntionsfourier}
\begin{align}
\begin{split}
    f^{ss}(c_s\eta p,c_s\eta|\mathbf{k}-\mathbf{p}|) &= \left (\frac{3+3w}{5+3w} \right )^2 \bigg ( \frac{2}{3(1+w)} \bigg [ \bigg ((T_{\Psi}(c_s\eta |\mathbf{k}-\mathbf{p}|) + \frac{T^{\prime}_{\Psi}(c_s\eta |\mathbf{k}-\mathbf{p}|)}{\mathcal{H}}\bigg )   \\
    &\times \bigg ( T_{\Psi}(c_s\eta p)+ \frac{T^{\prime}_{\Psi}(c_s\eta p)}{\mathcal{H}} \bigg )  \bigg ] 
    +T_{\Psi}(c_s\eta |\mathbf{k}-\mathbf{p}|)T_{\Psi}(c_s\eta p)\bigg )\text{ ,} 
\end{split}
\\
\begin{split}
     f^{st}(c_s \eta |\mathbf{k}-\mathbf{p}|,\eta p) &= \left (\frac{3+3w}{5+3w}\right )\bigg ( -2p^2T_{\Psi}(c_s \eta |\mathbf{k}-\mathbf{p}| ) T_{h}(\eta p)-2(\mathbf{k}-\mathbf{p})\cdot \mathbf{p} \\
    &\times T_{\Psi}(c_s \eta |\mathbf{k}-\mathbf{p}| ) T_{h}(\eta p)+ \mathcal{H}(1+3c_s^2)T_{\Psi}^{\prime } (c_s \eta |\mathbf{k}-\mathbf{p}| ) T_{h}(\eta p) \\
    &-(1-c_s^2)| \mathbf{k}-\mathbf{p}|^2T_{\Psi}(c_s \eta |\mathbf{k}-\mathbf{p}| ) T_{h}(\eta p)\bigg )\,,
\end{split}
\end{align}
\end{subequations}
and we also define the polarisation functions
\begin{subequations}\label{defpoldecomp}
\begin{align}
Q^{ss}_{\lambda}(\mathbf{k},\mathbf{p}) &= \left ( q^{ab}_{\lambda}(\mathbf{k}) \right)^*  p_{a}p_{b}\,, \\
Q^{st}_{\lambda , \lambda _1}(\mathbf{k},\mathbf{p}) &= \left( q^{ab}_{\lambda}(\mathbf{k}) \right)^*q^{\lambda _1}_{ab}(\mathbf{p})\,.
\end{align}
\end{subequations}
These obey some useful relationships (see App.~\ref{sec:Polarisation}) that simplify the computation of the GW spectrum. 

Having expressed the source terms in the form~\eqref{sourcetermsfourier}, we can write the solution to Eq.~(\ref{GW2eq}) for the GWs as
\begin{equation}  
    h^{(2)}_{\lambda}(\eta , \mathbf{k}) = \frac{4}{a(\eta)} \int ^{\eta}_0 \,\D\overline{\eta} \,G_{\mathbf{k}} (\eta ,\overline{\eta} )a(\overline{\eta}) \left ( \mathcal{S}^{ss}_{\lambda} (\overline{\eta} , \mathbf{k})+\mathcal{S}^{st}_{\lambda} (\overline{\eta} , \mathbf{k})\right )\,.
\end{equation}
where $G_{\mathbf{k}} (\eta ,\overline{\eta} )$ is a Green's function defined as the solution to the equation
\begin{equation}
\label{Greensequation}
    G^{''}_{\mathbf{k}}(\eta,\bar{\eta})+2\mathcal{H}G^{'}_{\mathbf{k}}(\eta,\bar{\eta})+k^2 G_{\mathbf{k}}(\eta,\bar{\eta})=\delta(\eta-\bar{\eta})\,.
\end{equation}
Its form depends on when the second order waves are sourced, i.e. it varies with the equation of state of the universe, and we leave it unspecified for the time being. Finally, it is useful to separate the scalar-scalar and scalar-tensor contributions and expand them in the form
\begin{align} \label{GW2sol}
    \begin{split}
         h^{(2)}_{\lambda}(\eta , \mathbf{k}) &= 4 \int \frac{\D^3\mathbf{p}}{(2\pi)^{\frac{3}{2}}} \bigg [Q^{ss}_{\lambda}(\mathbf{k},\mathbf{p}) I^{ss}(p,|\mathbf{k}-\mathbf{p}|,c_s,\eta)\mathcal{R}_{\mathbf{p}} \mathcal{R}_{\mathbf{k}-\mathbf{p}} + \sum_{\lambda _1} Q^{st}_{\lambda , \lambda _1}(\mathbf{k},\mathbf{p}) \\
         &\times I^{st}(p,|\mathbf{k}-\mathbf{p}|,c_s,\eta)h_{\mathbf{p}}^{\lambda _1} \mathcal{R}_{\mathbf{k}-\mathbf{p}}   \bigg] \text{ ,}
    \end{split}
\end{align}
where we have defined the two kernels
\begin{subequations}
\begin{align}
\label{kernelss}
I^{ss}(p,|\mathbf{k}-\mathbf{p}|,c_s,\eta)&= \int _{0}^{\eta} \,\D\overline{\eta}\,G_{\mathbf{k}} (\eta ,\overline{\eta} ) \frac{a(\overline{\eta})}{a(\eta)} f^{ss}(c_s\eta p,c_s\eta|\mathbf{k}-\mathbf{p}|) \,,
\\
\label{kernelst}
I^{st}(p,|\mathbf{k}-\mathbf{p}|,c_s,\eta)&= 
\int _{0}^{\eta} \,\D\overline{\eta}\,G_{\mathbf{k}} (\eta ,\overline{\eta})\frac{a(\overline{\eta})}{a(\eta)} f^{st}(c_s\eta |\mathbf{k}-\mathbf{p}|,\eta p) \, ,
\end{align}
\end{subequations}
which contain all the time dependence. Taking Eq.~\eqref{GW2sol} as our solution to the SOGW equation~\eqref{GW2eq}, we now employ it to compute the power spectrum associated with these waves.
\subsection{The power spectrum of scalar-scalar and scalar-tensor induced GWs}
We now consider the computation of the power spectrum of the induced SOGWs, $P_{\lambda}^{h^{(2)}}(\eta, k)$, via its decomposition into scalar-scalar and scalar-tensor disconnected and connected terms. The power spectrum is defined according to
\begin{equation}\label{2pointfunction}
    \langle h^{(2)}_{\lambda}(\eta, \mathbf{k})h^{(2)}_{\lambda ^{\prime}}(\eta, \mathbf{k^{\prime}}) \rangle =  \delta ^{(3)} (\mathbf{k} + \mathbf{k^{\prime}}) \delta ^{\lambda \lambda ^{\prime}}  P_{\lambda}^{h^{(2)}}(\eta, k)\,.
\end{equation}
Inserting the SOGW solution~\eqref{GW2sol} into Eq.~\eqref{2pointfunction}, we see that the computation of the power spectrum will, schematically, require the evaluation of the correlators
\begin{eqnarray}
\label{PSbreakup}
    \langle h^{(2)}_{\lambda}(\eta, \mathbf{k})h^{(2)}_{\lambda ^{\prime}}(\eta, \mathbf{k^{\prime}}) \rangle &\sim & \langle \mathcal{R}_{\mathbf{p}} \mathcal{R}_{\mathbf{k}-\mathbf{p}} \mathcal{R}_{\mathbf{p^{\prime}}} \mathcal{R}_{\mathbf{k^{\prime}}-\mathbf{p^{\prime}}} \rangle + \langle \mathcal{R}_{\mathbf{p}} \mathcal{R}_{\mathbf{k}-\mathbf{p}}h^{\lambda _1 ^{\prime}}_{\mathbf{p^{\prime}}}\mathcal{R}_{\mathbf{k^{\prime}}-\mathbf{p^{\prime}}}  \rangle  \nonumber \\
    &+& \langle h_{\mathbf{p}}^{\lambda _1} \mathcal{R}_{\mathbf{k}-\mathbf{p}} \mathcal{R}_{\mathbf{p^{\prime}}} \mathcal{R}_{\mathbf{k^{\prime}}-\mathbf{p^{\prime}}}  \rangle+\langle h_{\mathbf{p}}^{\lambda _1} \mathcal{R}_{\mathbf{k}-\mathbf{p}}h^{\lambda _1 ^{\prime}}_{\mathbf{p^{\prime}}}\mathcal{R}_{\mathbf{k^{\prime}}-\mathbf{p^{\prime}}}  \rangle\,.
\end{eqnarray}
At first order, perturbations have been separated into scalar and tensor modes. These first-order modes decouple and evolve independently, meaning that $\mathcal{R}_\mathbf{k}$ and $h_\mathbf{p}^\lambda$ are uncorrelated, and therefore we can set the two point function of a tensor and scalar mode to zero. Additionally, we assume $\langle \mathcal{R}_{\mathbf{k}} \rangle= \langle h^{\lambda}_{\mathbf{k}}\rangle=0$ for all modes, $\mathbf{k}$, and polarisations, $\lambda$, and that the fluctuations $h^{\lambda}_{\mathbf{k}}$ have Gaussian statistics\footnote{In this work we will only be concerned with the effects of scalar non-Gaussianity on the scalar-tensor contribution to the GW spectrum. We defer considerations of tensor non-Gaussianity to future work.}. Under these assumptions, the second and third terms of Eq.~\eqref{PSbreakup} vanish, whilst the last term is inherently disconnected (i.e. it can be broken down into a product of power spectra). The SOGW power spectrum then schematically reduces to
\begin{equation}
    \langle h^{(2)}_{\lambda}(\eta, \mathbf{k})h^{(2)}_{\lambda ^{\prime}}(\eta, \mathbf{k^{\prime}}) \rangle \sim \langle \mathcal{R}_{\mathbf{p}} \mathcal{R}_{\mathbf{k}-\mathbf{p}} \mathcal{R}_{\mathbf{p^{\prime}}} \mathcal{R}_{\mathbf{k^{\prime}}-\mathbf{p^{\prime}}} \rangle +\langle h_{\mathbf{p}}^{\lambda _1} \mathcal{R}_{\mathbf{k}-\mathbf{p}}h^{\lambda _1 ^{\prime}}_{\mathbf{p^{\prime}}}\mathcal{R}_{\mathbf{k^{\prime}}-\mathbf{p^{\prime}}}  \rangle _d\,.
\end{equation}
The four point scalar correlation contains both a disconnected and non-vanishing connected part\footnote{The subscript `c' stands for connected and the subscript `d' for disconnected.}
\begin{equation}
\label{condiscon}
    \langle \mathcal{R}_{\mathbf{p}} \mathcal{R}_{\mathbf{k}-\mathbf{p}} \mathcal{R}_{\mathbf{p^{\prime}}} \mathcal{R}_{\mathbf{k^{\prime}}-\mathbf{p^{\prime}}} \rangle = \langle \mathcal{R}_{\mathbf{p}} \mathcal{R}_{\mathbf{k}-\mathbf{p}} \mathcal{R}_{\mathbf{p^{\prime}}} \mathcal{R}_{\mathbf{k^{\prime}}-\mathbf{p^{\prime}}} \rangle _c + \langle \mathcal{R}_{\mathbf{p}} \mathcal{R}_{\mathbf{k}-\mathbf{p}} \mathcal{R}_{\mathbf{p^{\prime}}} \mathcal{R}_{\mathbf{k^{\prime}}-\mathbf{p^{\prime}}} \rangle _d \,,
\end{equation}
where the connected part is
\begin{equation}
    \langle \mathcal{R}_{\mathbf{p}} \mathcal{R}_{\mathbf{k}-\mathbf{p}} \mathcal{R}_{\mathbf{p^{\prime}}} \mathcal{R}_{\mathbf{k^{\prime}}-\mathbf{p^{\prime}}} \rangle _c = \delta ^{3}\left ( \mathbf{k}+\mathbf{k^{\prime}} \right ) \mathcal{T}\left (\mathbf{k},\mathbf{k}-\mathbf{p}, \mathbf{k^{\prime}},\mathbf{k^{\prime}}-\mathbf{p^{\prime}} \right ) \,,
\end{equation}
with $\mathcal{T}(\mathbf{k}_1,\mathbf{k}_2,\mathbf{k}_3,\mathbf{k}_4)$ the connected trispectrum. The disconnected part of Eq.~\eqref{condiscon} can be broken down further and simplified as
\begin{equation}
    \langle \mathcal{R}_{\mathbf{p}} \mathcal{R}_{\mathbf{k}-\mathbf{p}} \mathcal{R}_{\mathbf{p^{\prime}}} \mathcal{R}_{\mathbf{k^{\prime}}-\mathbf{p^{\prime}}} \rangle _d = \langle \mathcal{R}_{\mathbf{p}} \mathcal{R}_{\mathbf{p^{\prime}}} \rangle\langle \mathcal{R}_{\mathbf{k}-\mathbf{p}} \mathcal{R}_{\mathbf{k^{\prime}}-\mathbf{p^{\prime}}} \rangle + \langle \mathcal{R}_{\mathbf{p}} \mathcal{R}_{\mathbf{k^{\prime}}-\mathbf{p^{\prime}}}  \rangle \langle \mathcal{R}_{\mathbf{k}-\mathbf{p}} \mathcal{R}_{\mathbf{p^{\prime}}} \rangle \,.
\end{equation}
The scalar-tensor correlation reduces to
\begin{equation}
  \langle h_{\mathbf{p}}^{\lambda _1} \mathcal{R}_{\mathbf{k}-\mathbf{p}}h^{\lambda _1 ^{\prime}}_{\mathbf{p^{\prime}}}\mathcal{R}_{\mathbf{k^{\prime}}-\mathbf{p^{\prime}}}  \rangle _d = \langle h_{\mathbf{p}}^{\lambda _1} h_{\mathbf{p^{\prime}}}^{\lambda _1 ^{\prime}} \rangle \langle \mathcal{R}_{\mathbf{k}-\mathbf{p}} \mathcal{R}_{\mathbf{k^{\prime}}-\mathbf{p^{\prime}}} \rangle \,,
\end{equation}
hence we see that the power spectrum $P_{\lambda}^{h^{(2)}}$ defined in Eq.~(\ref{2pointfunction}) can be separated into three contributions
\begin{equation}
    P_{\lambda}^{h^{(2)}}(\eta , k) =  P_{\lambda}^{ss}(\eta , k)|_c + P_{\lambda}^{ss}(\eta , k)|_d + P_{\lambda}^{st}(\eta , k)|_d \,,
\end{equation}
each given by
\begin{subequations} \label{NGpowerspectrum}
\begin{align}
\begin{split}
     P_{\lambda}^{ss}(\eta , k)|_c&= 16  \int \frac{\D^3\mathbf{p}}{(2\pi)^{\frac{3}{2}}} \int \frac{d^3\mathbf{p^{\prime}}}{(2\pi)^{\frac{3}{2}}}  Q^{ss}_{\lambda}(\mathbf{k},\mathbf{p}) Q^{ss}_{\lambda}(\mathbf{k^{\prime}},\mathbf{p^{\prime}})  \\
    &\times I(p,|\mathbf{k}-\mathbf{p}|,c_s,\eta) I(p^{\prime},|\mathbf{k^{\prime}}-\mathbf{p^{\prime}}|,c_s,\eta)\mathcal{T}\left (\mathbf{k},\mathbf{k}-\mathbf{p}, \mathbf{k^{\prime}},\mathbf{k^{\prime}}-\mathbf{p^{\prime}} \right )  \,,
\end{split}
\\
\begin{split}
    P_{\lambda}^{ss}(\eta , k)|_d &=32 \int \frac{\D^3\mathbf{p}}{(2\pi)^{3}} Q^{ss}_{\lambda}(\mathbf{k},\mathbf{p}) Q^{ss}_{\lambda}(-\mathbf{k},-\mathbf{p})I(p,|\mathbf{k}-\mathbf{p}|,c_s,\eta)^2 \\
   &\times P_{\mathcal{R}}(p)P_{\mathcal{R}}(|\mathbf{k}-\mathbf{p}|)\,,
\end{split}
\\
\begin{split}\label{stpowerspectrum}
   P_{\lambda}^{st}(\eta , k)|_d &=16 \sum_{\lambda _1} \int \frac{\D^3\mathbf{p}}{(2\pi)^{3}} Q^{st}_{\lambda , \lambda _1}(\mathbf{k},\mathbf{p}) Q^{st}_{\lambda , \lambda _1}(-\mathbf{k},-\mathbf{p}) I(p,|\mathbf{k}-\mathbf{p}|,c_s,\eta)^2 \\
   &\times P_{h}^{\lambda _1}(|\mathbf{k}-\mathbf{p}|)P_{\mathcal{R}}(p)  \,.
\end{split} 
\end{align}
\end{subequations}
While we have assumed primordial tensor perturbations are Gaussian, we are yet to specify the statistics of the scalar curvature perturbation $\mathcal{R}_{\mathbf{k}}$.

\section{Inclusion of local-type scalar non-Gaussianity}\label{NGsection}
In this section we expand the scalar curvature perturbation to $\mathcal{O}(F_{\textrm{NL}})$ as a local-type non-Gaussian field. Substituting this into the expressions for the GW power spectrum derived in the previous section, we separate the contribution from the scalar-tensor interactions into a Gaussian and non-Gaussian part. The Gaussian part has been studied previously, see for example Refs.~\cite{Bari:2023rcw,Picard:2023sbz}, but the non-Gaussian part is new. We derive an expression for the contribution of this new term to the observable energy density, $\Omega(\eta_0,k)h^2$, ready for numerical evaluation in Sec.~\ref{resultsSec}.
\subsection{Primordial scalar non-Gaussianity in the scalar-tensor sector}
We model the primordial scalar non-Gaussianity as a local-type expansion of the scalar curvature perturbation field in real space
\begin{equation}
\label{localNGexpansion}
    \mathcal{R}(\mathbf{x}) = \mathcal{R}_\textrm{G}(\mathbf{x}) + F_{\textrm{NL}} \left (\mathcal{R}^2_\textrm{G}(\mathbf{x}) - \langle \mathcal{R}^2_\textrm{G}(\mathbf{x}) \rangle \right ) \, ,
\end{equation}
where $\mathcal{R}_\textrm{G}(\mathbf{x})$ is a scalar curvature perturbation with Gaussian statistics, and $F_{\textrm{NL}}$\footnote{This is related to the local non-Gaussianity parameter as $F_{\textrm{NL}}=\frac{3}{5}f_{\textrm{NL}}^{\textrm{local}}$.} is the local non-Gaussianity parameter. Other works have considered the effects of expanding the curvature perturbation to $\mathcal{O}(G_{\textrm{NL}})$ and beyond~\cite{Perna:2024ehx,Zeng:2024ovg,Pi:2024jwt,Li:2023xtl}, but we will restrict ourselves here to just $\mathcal{O}(F_{\textrm{NL}})$. Although $F_{\textrm{NL}}$ is weakly constrained on Planck scales~\cite{Planck:2019kim}, there is a lot of freedom in its magnitude on short scales. Indeed, since we will consider toy models which are peaked with large amplitudes on short scales, we expect that the magnitude of PNG may be considerably larger than CMB constraints on these scales. In Fourier space, the local-type expansion of $\mathcal{R}(\mathbf{x})$ shows up as a correction to the Gaussian power spectrum
\begin{equation}
\label{NGPS}
    P_{\mathcal{R}}(k) = P_{\mathcal{R},\textrm{G}}(k) + 2F_{\textrm{NL}}^2 \int \frac{\D^3\mathbf{p}}{(2\pi)^{3}}P_{\mathcal{R},\textrm{G}}(p)P_{\mathcal{R}, \textrm{G}}(|\mathbf{k}-\mathbf{p}|)  \,,
\end{equation}
with $ P_{\mathcal{R},\textrm{G}}(k)$ being the power spectrum of the Gaussian scalar perturbations from inflation. The effect of scalar non-Gaussianity on the scalar-tensor contribution to the GW spectrum can be seen by substituting the expansion of the power spectrum~\eqref{NGPS} into Eq.~\eqref{NGpowerspectrum}. This is the same procedure employed to study the effect of local non-Gaussianity on the SIGW spectrum (contributions from scalar-scalar interactions only), see e.g. Ref.~\cite{Adshead:2021hnm}. The terms resulting from this substitution in the scalar-scalar case were detailed in Ref~\cite{Adshead:2021hnm} to $\mathcal{O}(F_{\textrm{NL}}^4)$ and are reproduced in App.~\ref{NGSIGWS}. In the scalar-tensor sector, we see that the result of substituting the expansion of the power spectrum~\eqref{NGPS} into Eq.~(\ref{stpowerspectrum}) is a Gaussian contribution
\begin{align}\label{stG}
    \begin{split}
       P_{\lambda}^{st}(\eta , k)_{\textrm{Gaussian}}&=16 \sum_{\lambda _1} \int \frac{\D^3\mathbf{p}}{(2\pi)^{3}} Q^{st}_{\lambda , \lambda _1}(\mathbf{k},\mathbf{p}) Q^{st}_{\lambda , \lambda _1}(-\mathbf{k},-\mathbf{p})I^{st}(p,|\mathbf{k}-\mathbf{p}|,c_s,\eta)^2 \\
   &\times P_{h}^{\lambda _1}(|\mathbf{k}-\mathbf{p}|)P_{\mathcal{R},\textrm{G}}(p) \, , 
    \end{split}
\end{align}
and a new non-Gaussian scalar-tensor term
\begin{align} \label{stHS}
    \begin{split}
        P_{\lambda}^{st}(\eta , k)_{\textrm{HST}}&= 32 F_{\textrm{NL}}^2 \sum_{\lambda _1} \int \frac{\D^3\mathbf{p}}{(2\pi)^{3}}Q^{st}_{\lambda , \lambda _1}(\mathbf{k},\mathbf{p}) Q^{st}_{\lambda , \lambda _1}(-\mathbf{k},-\mathbf{p})I^{st}(p,|\mathbf{k}-\mathbf{p}|,c_s,\eta)^2 \\
        &\times  P_{h}^{\lambda _1}(p) \int \frac{\D^3\mathbf{q}}{(2\pi)^{3}} P_{\mathcal{R},\textrm{G}}(q)P_{\mathcal{R},\textrm{G}}(|\mathbf{k}-\mathbf{p}-\mathbf{q}|) \, .
    \end{split}
\end{align}
which we term the `hybrid scalar-tensor' (HST) contribution. The rest of this work will focus on the effects of this new term on the GW spectrum from scalar-tensor and combined scalar-scalar and scalar-tensor interactions.

\subsection{Expressions for the numerical evaluation of the scalar-tensor contributions}
\label{sec:changecoords}
The current form of the HST term~\eqref{stHS} is not amenable to numerical evaluation. In what follows, we derive the contribution of HST term to the spectral density of GWs that would be observed today, $\Omega(\eta_0,k)h^2$. For this, we need to evaluate the polarisation functions and switch to coordinates suitable for performing the integrals in Eq.~\eqref{stG} and Eq.~\eqref{stHS}.

We begin by defining the variables
\begin{equation}
\label{ucoord}
    u_1=\frac{|\mathbf{k}-\mathbf{p}|}{k}\,,
\end{equation}
\begin{equation}
\label{vcoord}
    v_1=\frac{p}{k}\,,
\end{equation}
along with the coordinate $\phi_p$ which is defined as the angle between the momenta $\mathbf{k}$ and $\mathbf{p}$. In these coordinates, the integral over the vector $\mathbf{p}$ can be recast as
\begin{equation}
    \int \,\D^3 \mathbf{p}=k^3 \int^\infty_0 \,\D v_1 \,\int^{1+v}_{|1-v|}\,\D u _1 \,  u_1v_1 \int^{2\pi}_0 \, \D \phi_p \,\, .
\end{equation}
To obtain a rectangular integration region, suitable for numerical evaluation, we also introduce the variables
\begin{equation}
    s_1=u_1-v_1 \, ,
\end{equation}
\begin{equation}
    t_1 = u_1+v_1+1 \, .
\end{equation}
The integration over the variables $u_1$ and $v_1$ can then be written as
\begin{equation}
    \int_0^\infty \, \D v_1 \, \int^{1+v_1}_{|1-v_1|} \,\D u_1 \, = \frac{1}{2} \int^{\infty}_0 \,\D t_1 \,\int^1_{-1} \,\D s_1 \, \,.
\end{equation}
For the Gaussian scalar-tensor term, we apply the preceding transformations and evaluate the polarisation functions, $Q^{st}_{\lambda, \lambda_1}$, in the new coordinates. See App.~\ref{sec:Polarisation} for details on the polarisation functions. The result is
\begin{align}\label{stGspherical}
    \begin{split}
       P_{R/L}^{st}(\eta , k)_\textrm{Gaussian}&= \frac{1}{32} \frac{k^3}{(2\pi)^2} \int ^{\infty}_0 \,\D t_1 \, \int ^{1}_{-1} \,\D s_1 \, \frac{u_1}{v_1^3} \bigg [ \left (u_1^2-(v_1+1)^2 \right )^4 P^{R/L}_{h}(kv_1) \\ 
       &+ \left (u_1^2-(v_1-1)^2 \right )^4 P^{L/R}_{h}(kv_1)  \bigg ]  \mathcal{I}_{st}(k,v_1,u_1,c_s)^2 P_{\mathcal{R},G}(ku_1)\,,
    \end{split}
\end{align}
where we have performed the integral over $\phi_p$ (there is no $\phi_p$ dependence in the Gaussian scalar-tensor term, so this integral simply evaluates to $2\pi$), and we have left the integrand written in the variables $v_1$ and $u_1$ for notational convenience.

For the non-Gaussian contribution~\eqref{stHS}, we need to introduce an extra set of variables
\begin{equation}
    u_2=\frac{|\mathbf{k}-\mathbf{p}-\mathbf{q}|}{|\mathbf{k}-\mathbf{p}|}\,,
\end{equation}
\begin{equation}
    v_2=\frac{q}{|\mathbf{k}-\mathbf{p}|}\,,
\end{equation}
where we now have two azimuthal angles $\phi_p$ and $\phi_q$. Starting from Eq.~\eqref{stHS}, we transform to the coordinates $(u_1,v_1,u_2,v_2, \phi_p,\phi_q)$ and integrate over the azimuthal angles immediately, since the HST term is disconnected and has no dependence on them. This generates a factor of $(2\pi)^2$. We then express the polarisation functions in terms of $u_1$ and $v_1$, as we did for the Gaussian scalar-tensor term, and transform to the coordinates more suitable for numerical integration
\begin{equation}
    s_2=u_2-v_2 \, ,
\end{equation}
\begin{equation}
    t_2 = u_2+v_2+1 \, .
\end{equation}
The result is the expression
\begin{align} \label{srHSspherical}
    \begin{split}
        P_{R/L}^{st}(\eta , k)_{\textrm{HST}}&= \frac{1}{32} F_{\textrm{NL}}^2 \frac{k^3}{(2\pi)^2} \int ^{\infty}_0 \,\D t_1\, \int ^{1}_{-1} \,\D s_1 \, \frac{u_1}{v_1^3} \bigg [ \left (u_1^2-(v_1+1)^2 \right )^4 P^{R/L}_{h}(kv_1) \\ 
        &+ \left (u_1^2-(v_1-1)^2 \right )^4 P^{L/R}_{h}(kv_1)  \bigg ] \mathcal{I}_{st}(k,v_1,u_1,c_s)^2 \\
        &\times  \frac{k^3}{(2\pi)^2} \int ^{\infty}_0 \,\D t_2 \int ^{1}_{-1} \, \D s_2 \, u_1^3v_2u_2 P_{\mathcal{R},\textrm{G}}(ku_1v_2)P_{\mathcal{R},\textrm{G}}(ku_1u_2) \, ,
    \end{split}
\end{align}
where the integrand is written in terms of the coordinates $(u_1,v_1,u_2,v_2)$ for simplicity.

The physical observable associated with GWs is the spectral density, $\Omega$, which is defined as \cite{Maggiore:1999vm}
\begin{equation}
    \Omega (\eta , k) = \frac{1}{12M_{\textrm{Pl}}^2} \frac{k^2}{\mathcal{H}^2(\eta)}\sum _{\lambda}  \overline{\mathcal{P}^{\lambda}_{h^{(2)}}(\eta ,k )} \, ,
\end{equation}
where the overline denotes a time average and $\mathcal{P}^{\lambda}_{h^{(2)}}(\eta ,k )$ is the dimensionless power spectrum of SOGWs\footnote{The dimensionless power spectrum is defined via \begin{equation*}
       \mathcal{P}^{\lambda}_{h^{(2)}} = \frac{k^3}{2\pi ^2}  P^{\lambda}_{h^{(2)}}\,.
\end{equation*}}. The present-day spectral density, $\Omega (\eta _0,k)$, is related to the spectral density at time of creation,  $\Omega (\eta ,k)$, by a dilution factor $\mathcal{N}=1.62\times10^{-5}$ such that
\begin{equation}
    \Omega (\eta _0,k)h^2 = \mathcal{N}\Omega (\eta ,k)\, ,
\end{equation}
where $h$ is the reduced Hubble's constant, $h=H_0/100$. $H_0$ is the present-day Hubble constant. To compute the contribution of scalar-tensor Gaussian and non-Gaussian terms to the present-day spectral density we therefore need to evaluate the time-averaged, polarisation-summed, dimensionless analogues of Eqs.~\eqref{stGspherical} and~\eqref{srHSspherical}. These are
\begin{align}
\label{Gaussian1}
    \begin{split}
       \overline{\mathcal{P}^{st}(\eta , k)_\textrm{Gaussian}}= \frac{1}{64}  &\int ^{\infty}_0 \,\D t_1 \, \int ^{1}_{-1} \,\D s_1 \, \frac{1}{u_1^2v_1^6} \big [ \left (u_1^2-(v_1+1)^2 \right )^4 + \left (u_1^2-(v_1-1)^2 \right )^4 \big] \\ 
       &\times \left ( \mathcal{P}^{R}_{h}(kv_1)  + \mathcal{P}^{L}_{h}(kv_1) \right ) \mathcal{P}_{\mathcal{R},G}(ku_1) \, \overline{\mathcal{I}_{st}(k,v_1,u_1,c_s)^2}\, ,
    \end{split}
\end{align}

\begin{align}
\label{HybridScalar1}
    \begin{split}
       \overline{\mathcal{P}^{st}(\eta , k)_{\textrm{HST}}}= \frac{F_{\textrm{NL}}^2}{128}  &\int ^{\infty}_0 \, \D t_1 \, \int ^{1}_{-1} \, \D s_1 \, \int ^{\infty}_0 \, \D t_2 \, \int ^{1}_{-1} \, \D s_2 \, \frac{1}{v_1^6u_1^2v_2^2u_2^2}\\
       &\times \left[ \left (u_1^2-(v_1+1)^2 \right )^4 + \left (u_1^2-(v_1-1)^2 \right )^4 \right] \\ 
       &\times \left ( \mathcal{P}^{R}_{h}(kv_1)  + \mathcal{P}^{L}_{h}(kv_1) \right ) \overline{\mathcal{I}_{st}(k,v_1,u_1,c_s)^2} \mathcal{P}_{\mathcal{R},\textrm{G}}(ku_1v_2)\mathcal{P}_{\mathcal{R},\textrm{G}}(ku_1u_2)  \,.
    \end{split}
\end{align}
Eq.~\eqref{HybridScalar1} is the new contribution to the induced GW spectrum from inflation that this work explores. In the next section we discuss its possible effects on the present-day spectral density and compare it to the other contributions already studied elsewhere in the literature.
\section{The effects of the non-Gaussian scalar-tensor term for a radiation dominated universe and peaked primordial power spectra} \label{resultsSec}
In this section we present numerically evaluated examples of induced SOGW spectra including Gaussian and non-Gaussian scalar and scalar-tensor contributions. Through this we may ascertain the impact of including non-Gaussian scalar-tensor contributions relative to the Gaussian and non-Gaussian scalar-scalar contributions that have already been studied. This will also allow us to highlight any pertinent inherent features of the non-Gaussian scalar-tensor term. Up until now, we have not assumed the equation of state of the universe, functional form of the primordial power spectra or the preservation of parity in the derivation of Eqs.~\eqref{Gaussian1} and~\eqref{HybridScalar1}. In fact, they are valid for any constant equation of state. From here we specialise to the concrete case of peaked scalar and tensor primordial power spectra in a radiation dominated (RD) universe, since this is the case most relevant for PBH production and the observation prospects of induced SOGWs. Strongly peaked sources are also the easiest cases to gain a semi-analytic handle on. We do this by specifying the transfer functions during RD, $T_\Psi$ and $T_h$, so that we may compute the time averaged kernel appearing in Eqs.~\eqref{Gaussian1} and~\eqref{HybridScalar1}, and the functional form of the scalar and tensor primordial power spectra.

The HST term~\eqref{HybridScalar1} is the `new' term that we will pay most attention to. We highlight the key features of the GW spectrum contribution it generates by combining numerical evaluations of the term for Gaussian-shaped primordial power spectra with semi-analytic results for the limiting case of a monochromatic spectrum. These features include: a sharp peak when the scalar and tensor primordial spectra peak at the same scale, a distinctive IR scaling, a cut-off of its UV tail and a divergence problem it shares with the Gaussian scalar-tensor contribution for primordial power spectra that are too wide~\cite{Bari:2023rcw}. We also compare the new term's GW signature to some of those studied already in the literature. To this end, in addition to computing the Gaussian and $\mathcal{O}(F_{\textrm{NL}}^2)$ scalar-tensor contributions to the induced SOGW spectrum, we also evaluate all terms to order $\mathcal{O}(F_{\textrm{NL}}^4)$ in the scalar-scalar sector. Expressions for these terms suitable for numerical evaluation and their Feynman diagrams are listed in App.~\ref{NGSIGWS} and~\ref{app:feynman} respectively. We do not detail their derivation any further in this work since they have been studied extensively already in Refs.~\cite{Adshead:2021hnm, Perna:2024ehx}. For simplicity, we also do not expand the curvature perturbation beyond what appears in Eq.~\eqref{localNGexpansion}, i.e. we don't introduce non-Gaussian parameters beyond $F_{\textrm{NL}}$ such as $G_{\textrm{NL}}$ or $H_{\textrm{NL}}$.

\subsection{Transfer functions and scalar-tensor kernel in radiation domination}
In a RD universe the equation of state takes a constant value $w=1/3$. With this fixed equation of state we can solve Eq.~\eqref{defsplitfluctuations} to obtain the form of the Green's function as
\begin{equation}
    G_{\mathbf{k}}(\eta,\bar{\eta})=\frac{ \sin{(k \eta - k \bar{\eta})}}{k} \, .
\end{equation}
Furthermore, the transfer functions for the scalar and tensor modes during RD can be determined from their definitions~\eqref{defsplitfluctuations} and the first-order Einstein field equations. These read 
\begin{subequations}\label{transferfunctionRD}
\begin{align}
T_{\Psi}(x)&= \frac{9}{x^2} \left (\frac{\sqrt{3}}{x}\sin{\frac{x}{\sqrt{3}}} - \cos{\frac{x}{\sqrt{3}}} \right )\, , \\
T_{h}(x)&=\frac{\sin{x}}{x} \, ,
\end{align}
\end{subequations}
where we have introduced the variable $x=k\eta$ and $c_s^2=w=1/3$. See e.g. Refs.~\cite{Adshead:2021hnm,Picard:2023sbz,Bari:2023rcw,Kohri:2018awv,Yu:2023lmo} for more details. This enables us to compute the time-averaged scalar-tensor kernel, $\overline{\mathcal{I}_{st}^2}$, appearing in Eqs.~\eqref{Gaussian1} and~\eqref{HybridScalar1}. Since we are interested in the spectrum at present times, $\Omega(\eta_0,k)h^2$, we take $\eta \rightarrow \infty$ (or $x\rightarrow \infty$) and evaluate the quantity $x^2\overline{\mathcal{I}_{st}^2(v,u)}$. This computation is tedious and detailed elsewhere in the literature, so we merely present the result 
\begin{align}
\label{kernelinRD}
    \begin{split}
        x^2\overline{\mathcal{I}_{st}^2(v,u)} &= \frac{1}{1152v^2u^6} \bigg [ \pi ^2 (u^2-3(v-1)^2)^2(u^2-3(v+1)^2)^2 \Theta (v+\frac{u}{\sqrt{3}}-1) + \bigg ( 4uv \\
        &\times \left (9-9v^2 +u^2 \right )-\sqrt{3}(u^2-3(v-1)^2)(u^2-3(v+1)^2) \log \bigg | \frac{\left (\sqrt{3}v-u \right )^2-3}{\left (\sqrt{3}v+u \right )^2-3} \bigg | \bigg )^2 \bigg ] \, ,
    \end{split}
\end{align}
and refer the reader to Refs.~\cite{Bari:2023rcw,Yu:2023lmo,Picard:2023sbz} for further details.
\begin{figure}[htbp]
\centering
\includegraphics[width=\linewidth]{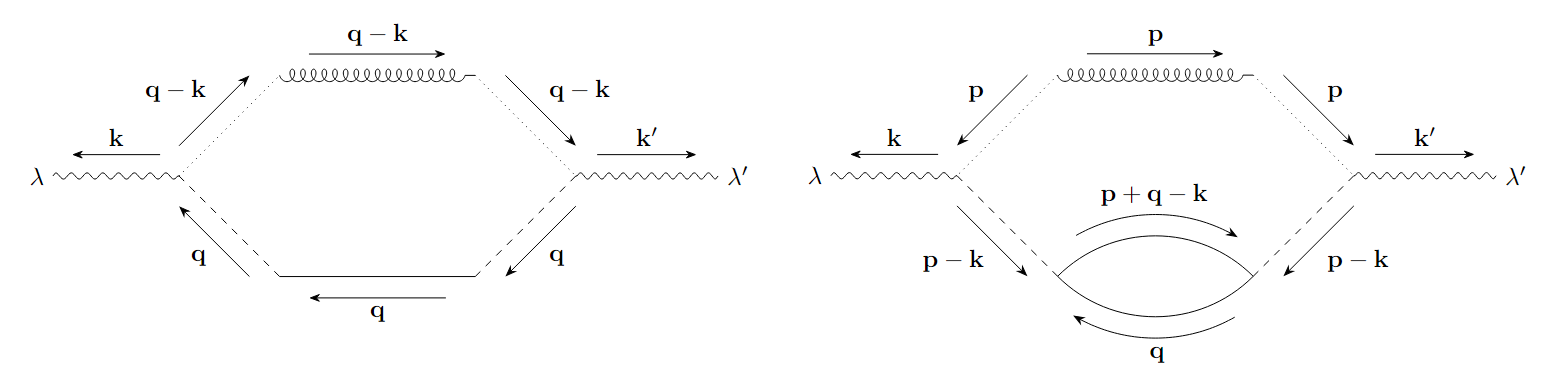}
\caption{\label{feynmandiagrams}
		Feynman diagrams for the scalar-tensor (left) Gaussian~\eqref{Gaussian1} and (right) hybrid scalar-tensor~\eqref{HybridScalar1} contributions to the induced SOGW spectrum. The `new' term explored in this work corresponds to the diagram on the right. The Feynman rules for these diagrams are listed in App.~\ref{app:feynman}.} 
\end{figure}
\subsection{UV cutoff of the HST term: monochromatic input power spectrum}
\label{UVcutoff}
Assuming the forms of the transfer functions~\eqref{transferfunctionRD} and scalar-tensor kernel~\eqref{kernelinRD} in RD, we can now compute the expression~\eqref{HybridScalar1} for the HST term by specifying the input Gaussian power spectra, $\mathcal{P}_{\mathcal{R},\textrm{G}}$ and $\mathcal{P}_{h,\textrm{G}}^{R/L}$. The behaviour of this contribution in the UV can be derived from the consideration of its form in the limiting case of monochromatic primordial scalar and tensor power spectra. 

We assume a monochromatic peak at the scale $k_p$ for both the scalar and tensor power spectra\footnote{Note that here we are assuming that the primordial spectra are peaking at the same scale. This need not be the case, however, see Refs. \cite{Bari:2023rcw,Picard:2023sbz} for setups where the input scalar and tensor spectra peak at different scales.}
\begin{equation}
    \mathcal{P}_{\mathcal{R},\textrm{G}}(k)=\mathcal{A}_s \delta\left(\ln\frac{k}{k_p}\right)\, \quad \textrm{and}\, \quad \mathcal{P}^{R/L}_{h}(k)=\mathcal{A}^{R/L}_t \delta\left(\ln\frac{k}{k_p}\right)\,,
\end{equation}
where $\mathcal{A}_s$ and $\mathcal{A}^{R/L}_t$ are the amplitudes of the scalar and tensor spectra respectively. For now, in the following analysis we will assume parity is not violated, i.e. $\mathcal{A}^{R}_t=\mathcal{A}^{L}_t=\mathcal{A}_t$. For this case, it has already been show in Refs.~\cite{Bari:2023rcw,Picard:2023sbz} that the Gaussian scalar-tensor contribution becomes
\begin{equation}
    \mathcal{P}^{st}(k)_{\textrm{Gaussian}}=\mathcal{A}_s \mathcal{A}_t\Tilde{k}^2\left ( 2+3\Tilde{k}^2+\frac{\Tilde{k}^4}{8} \right ) \overline{{\mathcal{I}^{st}}^2}_{u=v=\Tilde{k}^{-1}}\Theta(2-\Tilde{k})\,,
\end{equation}
where we have introduced the notation $\Tilde{k}=k/k_p$. This indicates the presence of a sharp UV cutoff in the contribution to the induced SOGW spectrum from this term; the Gaussian scalar-tensor term vanishes for $k\geq 2k_p$. While an analytic form for the Gaussian scalar-tensor contribution can be found for the case of monochromatic spectra, this is not the case for the non-Gaussian scalar-tensor terms. This is because, while in the Gaussian case we had two integrals over two Dirac-deltas, we now have four integrals over just three Dirac-deltas. It is still possible, however, to simplify the expression for the HST term by performing three integrals and then to study the behaviour of what remains. 

Starting with Eq.~\eqref{HybridScalar1} we see that, for monochromatic input spectra, the integrand will contain a product of three Dirac-deltas
\begin{equation}
    \delta\left(\ln\frac{k(1-s_1+t_1)}{2k_p}\right)\delta\left(\ln\frac{k(1+s_1+t_1)(1-s_2+t_2)}{4k_p}\right)\delta\left(\ln\frac{k(1+s_1+t_1)(1+s_2+t_2)}{4k_p}\right)\,.
\end{equation}
By integrating over the variables $t_1,t_2,s_2$, and keeping in mind the limits of the integration, it can be show that the contribution acquires an overall factor of a product of Heaviside functions
\begin{equation}
\label{GSTcutoff}
    \Theta\left(\frac{k_p-k s_1}{k_p+k s_1}\right)\Theta\left(-1+\frac{2k_p}{k}+s_1\right)\,.
\end{equation}
We can find the largest value of $k$ at which this combination is non-zero by setting the arguments of each Heaviside function to zero and solving for $k$
\begin{equation}
    k=\frac{2k_p}{1-s_1}\, \quad \textrm{or}\, \quad k=\frac{k_p}{s_1}\,.
\end{equation}
The maximum value of $k$ will occur for the value of $s_1$ at which these equations are simultaneously satisfied. This is when $s_1=1/3$. For this value of $s_1$, the Heaviside functions both become zero at $k= 3k_p$. This means that the scalar-tensor non-Gaussian term, the HST, will vanish for $k \geq 3k_p$. This is a different UV cutoff to the Gaussian scalar-tensor term~\eqref{GSTcutoff}, but coincides with the non-Gaussian scalar-scalar contributions' cutoffs~\cite{Adshead:2021hnm}. For broader input power spectra, the contribution to the spectrum from these terms will typically extend past their limiting-case UV cutoffs, but maintain their hierarchy. Terms with a UV cutoff at larger momenta in the monochromatic case will dominate at higher $k$-values over terms with a smaller UV cutoff. 

\subsection{Features of the HST contribution: Gaussian peaked spectra}
We now consider the more physical case of peaked primordial scalar and tensor power spectra with non-zero width. As a concrete example, we work with Gaussian spectra of the form
\begin{align}
    \mathcal{P}_{\mathcal{R},\textrm{G}}(k)&=\mathcal{A}_s\left(\frac{k}{k_s}\right)^3\frac{1}{\sqrt{2\pi (\sigma_s /k_s)^2}} \exp{\left(-\frac{(k-k_s)^2}{2\sigma_s^2}\right)}\,, \\
    \mathcal{P}^{R/L}_{h}(k)&=\mathcal{A}^{R/L}_t\left(\frac{k}{k_t}\right)^3\frac{1}{\sqrt{2\pi (\sigma_t /k_t)^2}} \exp{\left(-\frac{(k-k_t)^2}{2\sigma_t^2}\right)}\,,
\end{align}
where $\mathcal{A}_{\textrm{s/t}}$ are the peak amplitudes, $\sigma_{\textrm{s/t}}$ are the widths and $k_{\textrm{s/t}}$ are the positions of the peaks of the primordial scalar and tensor power spectra respectively. The power spectra are normalised such that $\int \D \ln k \, \mathcal{P}_{\mathcal{R},\textrm{G}}(k)=\mathcal{A}_s$ and $\int \D \ln k \, \mathcal{P}^{R/L}_{h}(k)=\mathcal{A}^{R/L}_t$. In the following, we typically position the scalar and tensor peaks at a scale accessible to LISA, corresponding to a physical frequency of $f_{\textrm{LISA}}=3.4$ mHz~\cite{Sathyaprakash:2009xs}. Furthermore, we typically set the amplitude of the scalar power spectrum to around $\mathcal{A}_s=10^{-2}$, since this is an important order of magnitude of the scalar power spectrum for both PBH production and generation of observable SIGW signals, and the tensor power spectrum to $\mathcal{A}^R_t=\mathcal{A}^L_t=0.1 \mathcal{A}_s$. This assumes that there is no parity violation and that tensor perturbations are generically smaller than scalar ones.

\begin{figure}
    \centering
    \includegraphics[width=\linewidth]{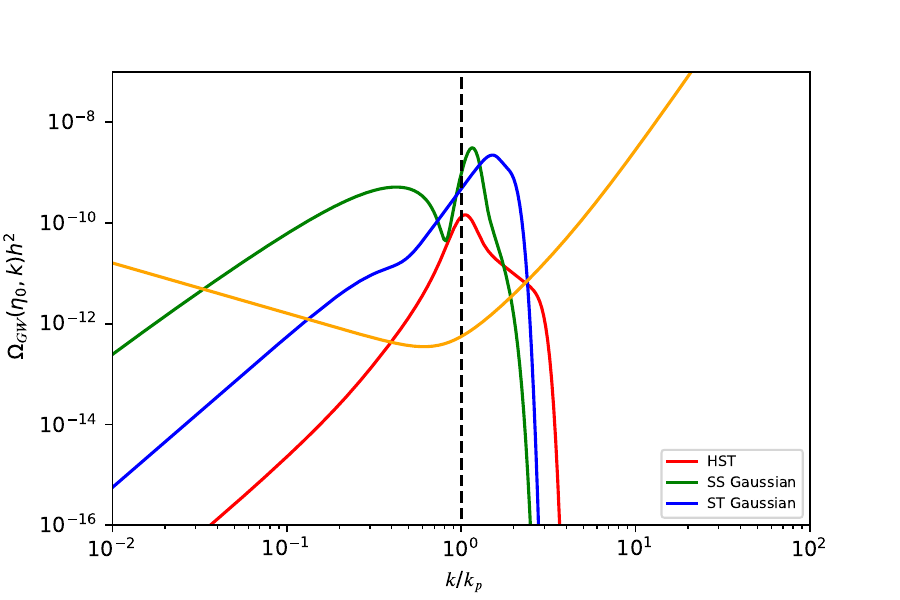}
    \caption{The present-day spectral density $\Omega_{\textrm{GW}}(\eta_0,k)h^2$ of SOGWs induced by scalar-scalar (green) and scalar-tensor (blue) modes at Gaussian order and scalar-tensor modes at $\mathcal{O}(F^2_{\textrm{NL}})$ (red) for Gaussian primordial scalar and tensor power spectra with amplitude $\mathcal{A}_s=10^{-2}$, $\mathcal{A}^{R/L}_t=0.1 \mathcal{A}_s$, with peak at $k=k_p$ and width $\sigma_s=\sigma_t=k_p/10$. The solid orange curve denotes the LISA sensitivity~\cite{Sathyaprakash:2009xs} and the dashed black line marks the location of the peaks in the scalar and tensor power spectra $k_p$. The non-Gaussianity parameter is set to $F_{\textrm{NL}}=1$.   }
    \label{fig:HybridScalar}
\end{figure}

In Fig.~\ref{fig:HybridScalar} we plot the contributions from the Gaussian scalar-scalar, Gaussian scalar-tensor and HST terms in the default scenario just described. This gives us an impression of the sort of effect introducing scalar non-Gaussianity might have on the induced SOGW spectrum through the scalar-tensor sector. Immediately, we notice that the HST term peaks at a scale closer to $k_p$ than either of the Gaussian contributions. It was shown previously that, for a delta-function primordial spectrum, the Gaussian scalar-scalar term peaks at $k=2/\sqrt{3}k_p$~\cite{Ananda:2006af}. We have verified numerically that, in the limit of monochromatic spectra, our new term peaks at $k=k_p$. This behaviour is unique to the HST term, other contributions from the scalar-scalar and scalar-tensor sectors generally peak at $k$-values after $k=k_p$. The scalar-tensor Gaussian peaks just a little before $k=(1+\frac{1}{\sqrt{3}})k_p$, which is the point at which the Heaviside function in the kernel~\eqref{kernelinRD} cuts out. Non-Gaussian scalar-scalar contributions typically peak either at the peak of the Gaussian, or at later `knees' around $k=2k_p$, which is the cutoff of the Gaussian contributions. The fact that the HST term peaks close to $k_p$ contributes further to the typical influence of non-Gaussianity on SOGW spectra --- the smoothing out of peaks in the SOGW spectrum. 

The additional smoothing effect of the new HST term is unlikely to be distinguishable from that of other non-Gaussian terms. A more distinctive feature of the HST term, however, is its large amplitude at shorter scales. At scales just beyond $k_p$, the scalar-tensor Gaussian term dominates over the scalar-scalar, but for scales even smaller than this eventually the HST term takes over. This is a result of the conservation of momentum discussed in the previous section~\ref{UVcutoff}. In the limit of monochromatic primordial spectra, the Gaussian terms have a UV cutoff at $k=2k_p$, whereas the HST contribution does not vanish until later at $k=3k_p$. When we consider Gaussian-shaped primordial spectra with finite width, the Gaussian and HST terms may extend a little beyond these cutoffs, but the dominance of the HST term at these scales is preserved.

This suggests that the distinctive feature of scalar non-Gaussianity in the scalar-tensor sector will be the presence of so-called `knees' at short scales in the spectrum, just like in the case of scalar-scalar non-Gaussianity. This presents us with two questions: is it plausible that LISA could detect such a knee, and can the knee from the HST term be distinguished from the knee generated by scalar-scalar non-Gaussian terms? In the following section we analyse the effect of the HST term on an induced SOGW spectrum when all scalar-scalar non-Gaussian contributions up to $\mathcal{O}(F^4_{\textrm{NL}})$ are also included. 

We also notice that the HST term has a similar infra-red running as the Gaussian scalar-tensor term in the limit $k\rightarrow0$. It was shown in Ref.~\cite{Bari:2023rcw} that the Gaussian scalar-tensor contribution lacks the logarithmic running associated with SIGWs. This is also true for the non-Gaussian scalar-tensor term. Interestingly though, although the HST tends to the same scaling as the Gaussian scalar-tensor in the far IR, it does so with an unusual concavity. 
\subsection{Comparison with scalar-scalar non-Gaussianity}
\label{sec:compare}
\begin{figure}
\centering
\begin{subfigure}{.49\textwidth}
  \centering
  \includegraphics[width=\linewidth]{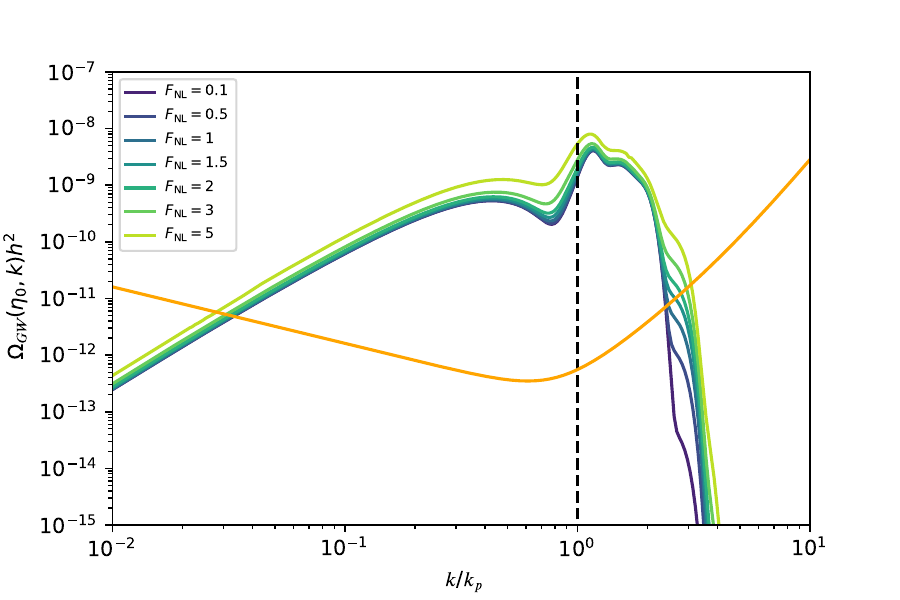}
\end{subfigure}
\begin{subfigure}{.49\textwidth}
  \centering
  \includegraphics[width=\linewidth]{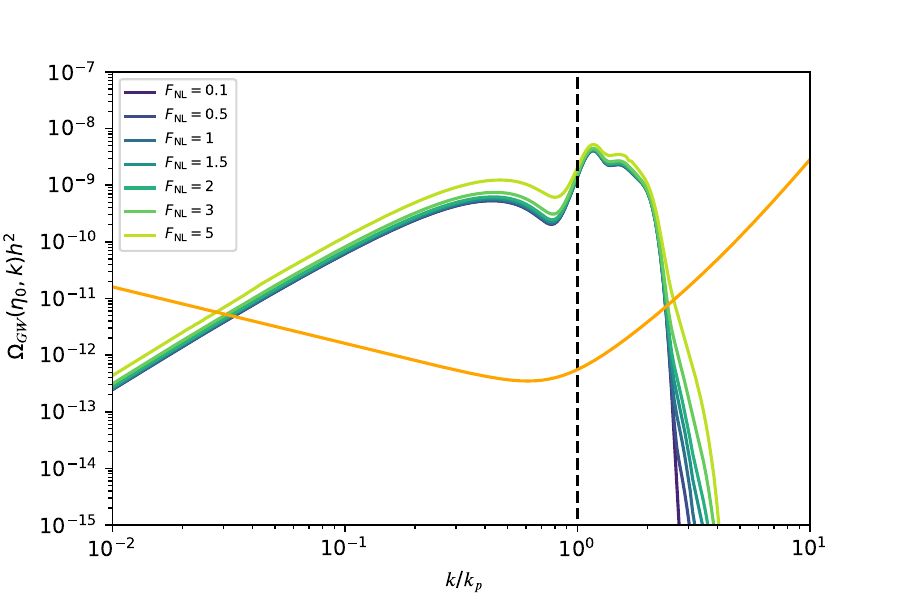}
\end{subfigure}
\begin{subfigure}{.49\textwidth}
  \centering
  \includegraphics[width=\linewidth]{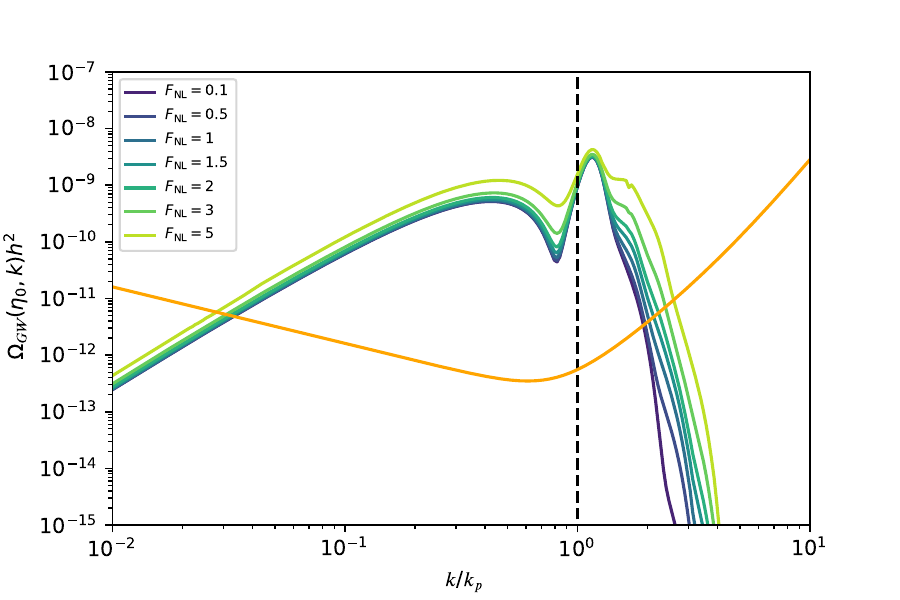}
\end{subfigure}
\caption{The present-day spectral density $\Omega_{\textrm{GW}}(\eta_0,k)h^2$ of SOGWs induced by scalar-scalar and scalar-tensor modes for varying values of $F_{\textrm{NL}}$ and Gaussian scalar and tensor power spectra centered at $k_p$ with width $1/10$ and $\mathcal{A}^{R/L}_t=0.1\mathcal{A}_s$. The solid orange line represents the LISA sensitivity band. The vertical black dashed line marks the scale $k_p$. (Top left) All scalar-scalar Gaussian and non-Gaussian contributions to order $\mathcal{O}(F^4_{\textrm{NL}})$ as well as the scalar-tensor Gaussian and HST terms are included. (Top right) Same as the previous, but the HST term has been excluded. (Bottom) same as the previous, but with the scalar-tensor Gaussian removed in addition.}
\label{fig:SS-STNGcompare}
\end{figure}
The key question we wish to address is whether the inclusion of scalar non-Gaussianity in the scalar-tensor sector can have a noticeable effect on the overall induced SOGW spectrum beyond the contributions from scalar-scalar Gaussian and non-Gaussian terms studied previously in the literature. To this end, in Fig.~\ref{fig:SS-STNGcompare} we plot the present-day spectral density $\Omega_{\textrm{GW}}(\eta_0,k)h^2$ of SOGWs with varying $F_{\textrm{NL}}$ parameters for three different cases:
\begin{enumerate}
    \item Including all Gaussian and non-Gaussian contributions, to $\mathcal{O}(F^4_{\textrm{NL}})$, from both scalar-scalar and scalar-tensor sectors
    \item Excluding just the non-Gaussian HST contribution
    \item Excluding the Gaussian and non-Gaussian scalar-tensor sector contributions
\end{enumerate}
This allows us to gauge the effect of the scalar-tensor sector on the overall spectrum, as well as to determine whether non-Gaussian scalar-tensor effects are distinguishable from all the other contributions.

Firstly, we point out that the existence of a Gaussian scalar-tensor contribution can be masked almost entirely by the non-Gaussian contributions from the scalar-scalar sector, but only for fairly large values of $F_{\textrm{NL}}$, i.e. $F_{\textrm{NL}}\geq 5$. The distinctive feature of the Gaussian scalar-tensor contribution, relative to the Gaussian scalar-scalar, is the continuation of the spectral density peak beyond $k=2k_p$. The amplitude of the Gaussian scalar-scalar term declines sharply around $k=2k_p$, whereas the Gaussian scalar-tensor term maintains a large amplitude until almost $k=3k_p$.
Regardless of the value of $F_{\textrm{NL}}$, if the scalar-tensor sector is significant, the amplitude does not fall off rapidly until past this threshold. If, on the other hand, the primordial tensor power spectrum does not have a large amplitude peak (i.e. if the scalar-tensor sector becomes negligible), then the scalar-scalar terms will dominate and the spectral density tends to drop substantially before the threshold. This is not the case, though, if $F_{\textrm{NL}}$ begins to approach $\mathcal{O}(10)$. Already at $F_{\textrm{NL}}=5$, we can see that the scalar-scalar sector alone can mimic the behaviour of the scalar-tensor Gaussian contribution, raising the power at scales around $k=2k_p$ to an amplitude comparable to the peak.

Turning to the non-Gaussian scalar-tensor contribution, we notice that it doesn't have any significant impact on the peak of the induced SOGW spectrum. This would suggest that the presence of a scalar-tensor sector is unlikely to be determined by LISA observations. Even if LISA detected an extended peak in the spectral density, whether it was caused by a scalar-tensor contribution or purely scalar non-Gaussianity would be uncertain. There is, however, one clear difference between the spectrum including scalar-tensor non-Gaussian terms and the others. In the top-left panel of Fig.~\ref{fig:SS-STNGcompare} the HST term clearly introduces an additional `kick' for scales $3k_p < k < 4k_p$. If $F_{\textrm{NL}}$ is of $\mathcal{O}(1)$ or larger, then the spectrum's terminal drop off in amplitude is delayed until almost $k=4k_p$. It is possible that this feature could uniquely signal the presence of a significant scalar-tensor sector, or even sizeable scalar non-Gaussianity. For our choice of $k_p$, it would be difficult for LISA to spot this extension for $F_{\textrm{NL}}$ values smaller than about 5, since the kick occurs right at the bottom of the LISA sensitivity. But, if $k_p$ is suitably smaller than the LISA scale $k_{\textrm{LISA}}$, then more of the drop-off in amplitude may be seen. In this case, $F_{\textrm{NL}}$ values as low as unity may produce kicks that LISA could detect.
\subsection{Varying primordial power spectrum widths}
\begin{figure}
\centering
\begin{subfigure}{.49\textwidth}
  \centering
  \includegraphics[width=\linewidth]{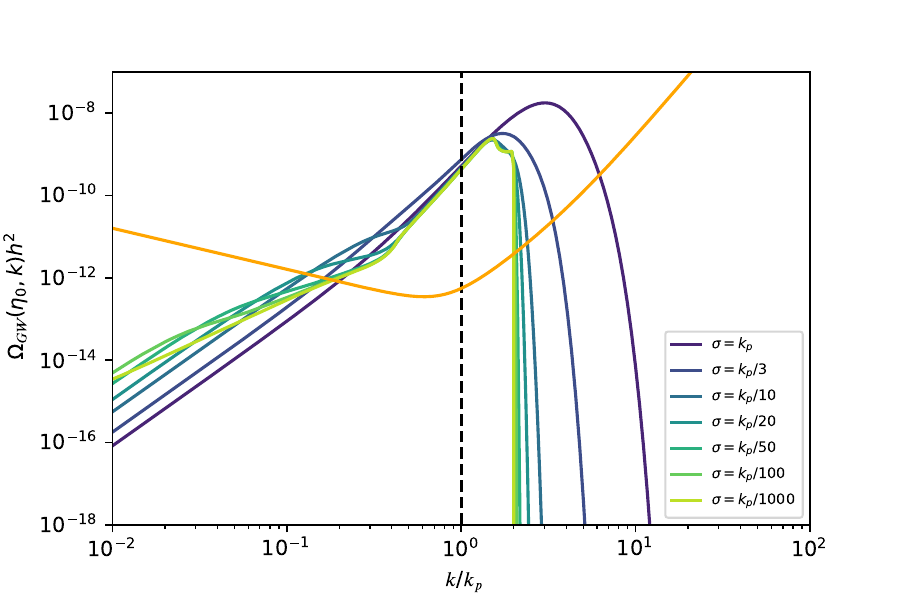}
\end{subfigure}
\begin{subfigure}{.49\textwidth}
  \centering
  \includegraphics[width=\linewidth]{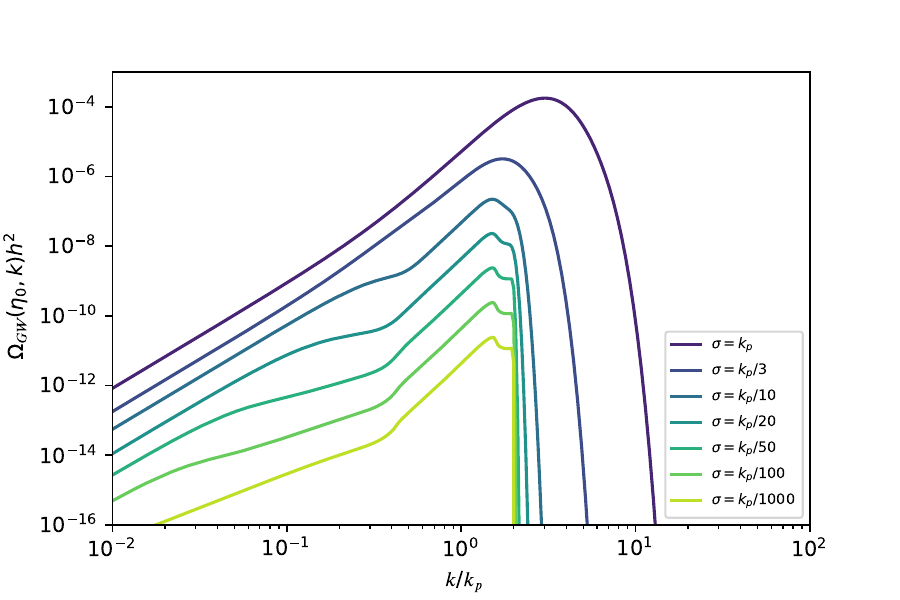}
\end{subfigure}
\begin{subfigure}{.49\textwidth}
  \centering
  \includegraphics[width=\linewidth]{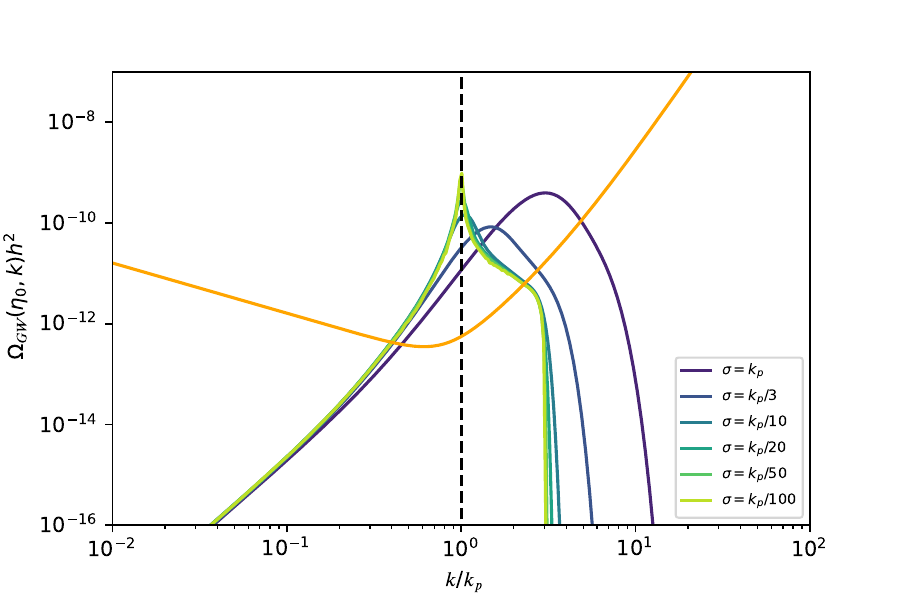}
\end{subfigure}
\begin{subfigure}{.49\textwidth}
  \centering
  \includegraphics[width=\linewidth]{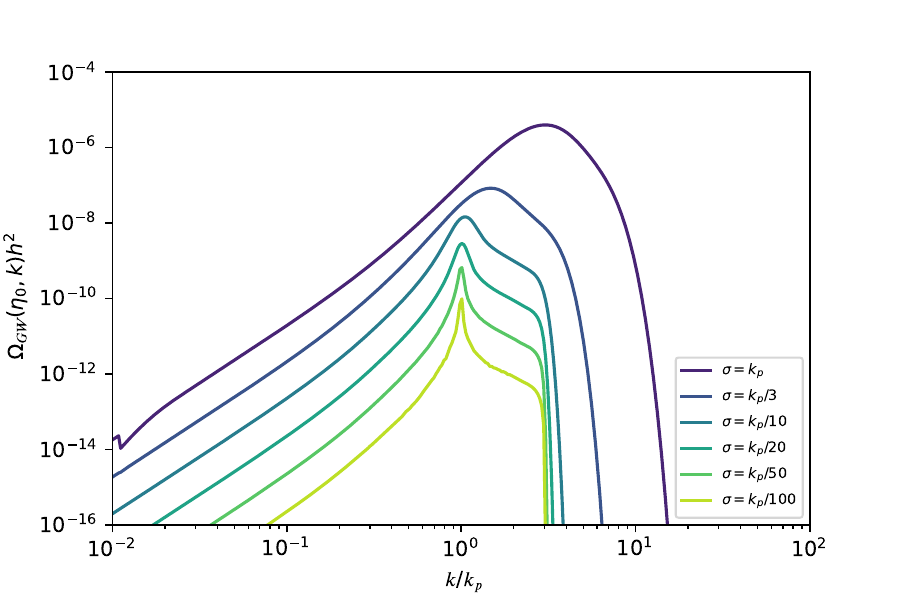}
\end{subfigure}
\caption{The contribution to the present-day spectral density, $\Omega_{\textrm{GW}}(\eta_0,k)h^2$, of SOGWs induced by the scalar-tensor Gaussian (top) and non-Gaussian HST (bottom) terms. On the left we plot these contributions for different values of the scalar and tensor power spectrum widths, $\sigma_s=\sigma_t=\sigma$, where $F_{\textrm{NL}}=1$ and $\mathcal{A}_s=10^{-2}$, $\mathcal{A}^{R/L}_t=0.1\mathcal{A}_s$. On the right we plot the same, but each curve, labelled $i$ with $0\leq i \leq5$, is multiplied by a factor $10^{-2+i}$. This is in order to space the curves out and enable an easier comparison of their differing features. The black dashed line marks the peak scale, $k_p$, and the solid orange curve represents the LISA sensitivity curve.}
\label{fig:STWidths}
\end{figure}
In Ref.~\cite{Bari:2023rcw} it was shown that the scalar-tensor induced SOGWs suffer from an unphysical enhancement, or a classical divergence, at Gaussian order when the input power spectra is broad. As $k\rightarrow \infty$, the integrand in Eq.~\eqref{Gaussian1} will begin to diverge, meaning that integrating over an insufficiently peaked spectra means integrating over theses `problematic' modes. The result is an unphysical enhancement of the induced SOGW observable spectral density. In Fig.~\ref{fig:STWidths} we plot the scalar-tensor Gaussian and HST contributions for a typical Gaussian scalar and tensor power spectrum, but with varying widths. Below, we comment on a few salient features.

First of all, we notice that narrower primordial power spectrum peaks lead to an increasingly sharp peak in the HST term at $k=k_p$. The height of this feature grows inversely with the width, but remains finite even in the limit of monochromatic scalar and tensor power spectra. In this regard, the HST term does not suffer from a divergence. We note, however, that the model appears to break down when the width of the primordial input spectra increases beyond about $\sigma=k_p/10$. When this occurs, the scalar-tensor Gaussian and HST contributions both deform away from their characteristic shape. The UV tails extend well beyond the expected cutoff, the peak of the spectrum shifts away from $k=k_p$ and its amplitude grows. These problems seem to occur after the same threshold widths for both the Gaussian and non-Gaussian scalar-tensor terms. 

Additionally, the width of the input primordial spectra has almost no effect on the IR tail of the spectral density for the HST contribution, but does affect the Gaussian. The IR tails of the Gaussian contributions have three distinct scalings: a $k^3$ growth on very large scales (i.e. in the IR limit), a slower intermediate growth of around $k^{3/2}$ and a final steep growth of roughly $k^5$ towards the peak. Changing the width of the scalar and tensor spectra affects how elongated the intermediate growth is.

\subsection{Primordial parity violation}
Certain models of inflation are capable of generating degrees of parity violation in the primordial tensor perturbations~\cite{Obata:2016tmo,Bartolo:2018elp,Bartolo:2017szm}. While primordial parity violation would have no effect on the spectrum of SIGWs (since they involve the interactions of primordial scalar perturbations only), it may be transmitted to SOGWs induced by scalar-tensor interactions through the affected primordial tensor perturbations. The effects of parity violation on the Gaussian scalar-tensor contribution were investigated in Ref.~\cite{Bari:2023rcw}. In this section we give a brief review of their results and take a step forward to consider the effects of parity violation on the non-Gaussian HST term. For simplicity we restrict our analysis to the right-handed spectral density of the scalar-tensor waves. This is an arbitrary choice and the same analysis could be performed for the left-handed case.  

\begin{figure}
    \centering
    \includegraphics[width=\linewidth]{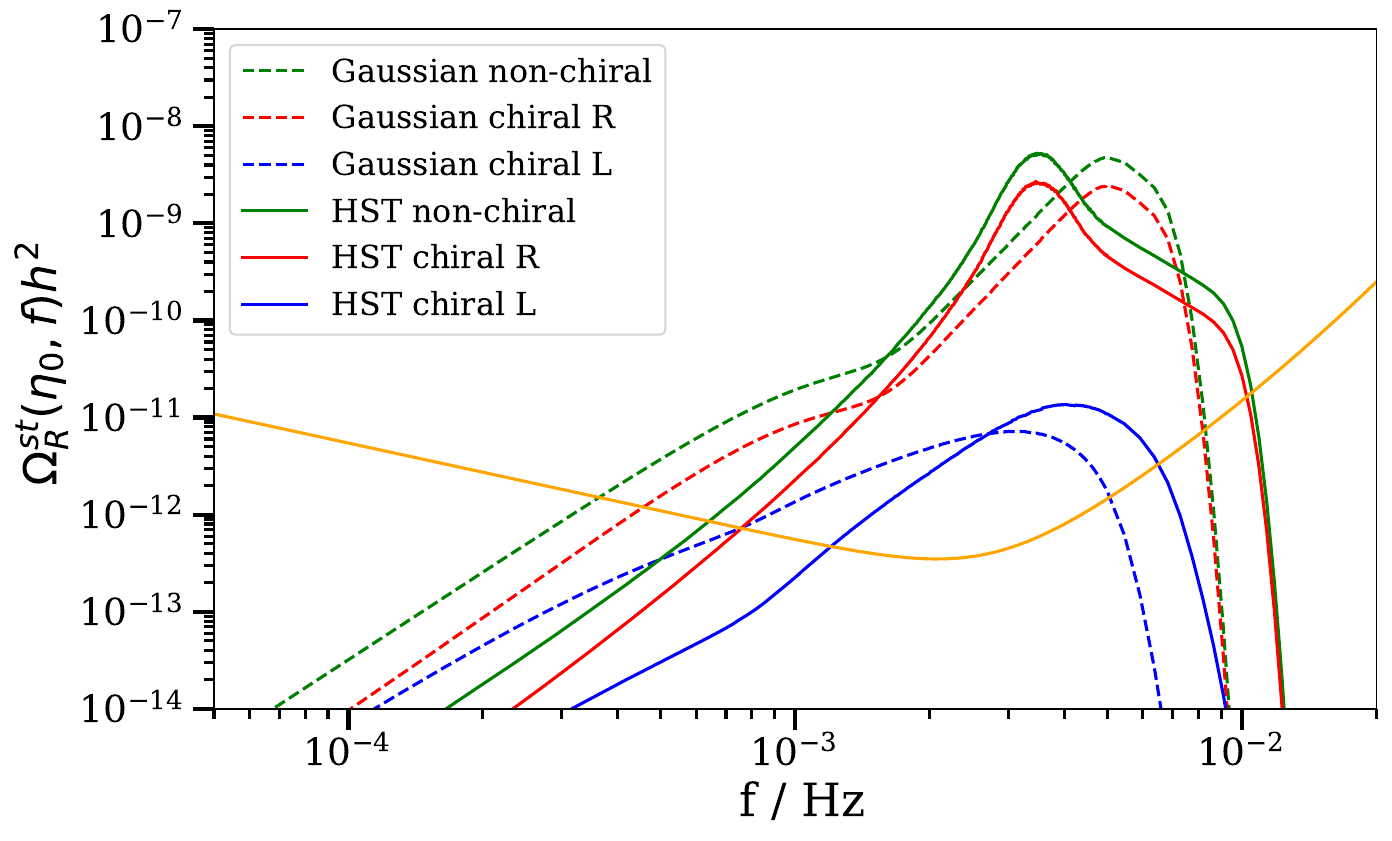}
    \caption{The present-day right-handed spectral density of scalar-tensor induced SOGWs, $\Omega^{st}_{R}(\eta_0,f)h^2$, as a function of frequency, $f$. Dashed lines corresponds to the Gaussian scalar-tensor contribution, solid lines to the non-Gaussian HST contribution. In green we have the non-chiral contributions, in red the contribution from only right-handed primordial tensor perturbations and in blue the contribution from only left-handed primordial tensor perturbations. We have fixed $F_{\textrm{NL}}=3$. The solid orange line is the projected LISA sensitivity curve.}
    \label{fig:chiralityplot}
\end{figure}

The right-handed spectral density of the Gaussian scalar-tensor contribution can be found straightforwardly using Eq.~\eqref{stGspherical}
\begin{align}
    \begin{split}
        \Omega _R^{st}(\eta , k)_\textrm{Gaussian} &= \frac{1}{768} \int _0^{\infty}{\rm d} t_1 \int _{-1}^{1}{\rm d} s_1 \, \frac{1}{v_1^6u_1^2} \bigg [ \left (u_1^2-(v_1+1)^2 \right )^4 \mathcal{P}^{R}_{h,g}(kv_1) + \left (u_1^2-(v_1-1)^2 \right )^4 \\
        &\times \mathcal{P}^{L}_{h,g}(kv_1)  \bigg ]\mathcal{P}_{\mathcal{R},G}(ku_1) x^2\overline{\mathcal{I}_{st}(k,v_1,u_1,c_s)^2}\,. 
    \end{split}
\end{align}
In Fig.~\ref{fig:chiralityplot} we consider what the right-handed spectral density looks like in the case of having either no primordial left-handedness (dashed red) or no primordial right-handedness (dashed blue). For comparison, we also include the non-chiral (i.e. no initial parity violation, so Eq.~\eqref{Gaussian1}) case (dashed green). As explained in Ref.~\cite{Bari:2023rcw}, the IR behaviour of the spectral density for the case of having an initial right- or left-handed parity is similar. On shorter scales, however, the spectral density corresponding to the no left-handedness scenario exhibits a large-amplitude peak, in contrast to its no right-handedness counterpart. 

We can extend this analysis to the HST term in Eq.~\eqref{HybridScalar1}, and derive an expression for the right-handed spectral density it generates
\begin{align}
    \begin{split}
        \Omega _R^{st}(\eta , k)_{HST} &= \frac{1}{1536} \int _0^{\infty}{\rm d} t_1 \int _{-1}^{1}{\rm d} s_1 \int _0^{\infty}{\rm d} t_2 \int _{-1}^{1}{\rm d} s_2 \, \frac{1}{v_1^6u_1^2v_2^2u_2^2} \bigg [ \left (u_1^2-(v_1+1)^2 \right )^4   \\
        &\times \mathcal{P}^{R}_{h,g}(kv_1) + \left (u_1^2-(v_1-1)^2 \right )^4 \mathcal{P}^{L}_{h,g}(kv_1)  \bigg ]x^2\overline{\mathcal{I}_{st}(k,v_1,u_1,c_s)^2} \\
        &\times \mathcal{P}_{\mathcal{R},G}(ku_1v_2)\mathcal{P}_{\mathcal{R},G}(ku_1u_2) \text{ .} 
    \end{split}
\end{align}
The contributions from this term for different primordial parity scenarios are displayed as solid lines in Fig.~\ref{fig:chiralityplot}. One can observe in Fig.~\ref{fig:chiralityplot} that the conclusions drawn for the Gaussian case in Ref.~\cite{Bari:2023rcw} and above remain applicable to the HST contribution. This fact is consistent with expectations, since any effects of primordial parity violation will be transmitted to scalar-tensor induced SOGWs through the tensor perturbations and our inclusion of scalar non-Gaussianity has not affected these.

\section{Discussion}
\label{sec:conclusions}
SOGWs induced by primordial scalar and tensor fluctuations are an exciting potentially observable component of the SGWB. Any detection of induced SOGWs by LISA, or other future GW observatories, would offer a wealth of information concerning inflationary dynamics at scales far smaller than the CMB. Specifically, since any observable induced SOGWs would require a significant amplification of either the scalar or tensor primordial power spectra on small scales, their detection could suggest departures from standard single-field slow-roll inflation, multi-field inflationary scenarios, non-Gaussian perturbations or the existence of PBHs. Many works have already investigated signatures of scalar-scalar and scalar-tensor induced SOGWs and explored the effects of scalar local non-Gaussianity on the scalar-scalar sector. In this paper we extend this analysis to study the effects of local scalar non-Gaussianity on the scalar-tensor induced SOGWs for the first time.

We find that the inclusion of local scalar non-Gaussianity leads to the appearance of a new term in the scalar-tensor sector --- the hybrid scalar-tensor. This is a disconnected contribution appearing at $\mathcal{O}(F_{\textrm{NL}}^2)$ in the local non-Gaussian expansion. In the typical RD scenario with coincident peaked scalar and tensor primordial power spectra, the new term shares many similarities with the Gaussian scalar-tensor and non-Gaussian scalar-scalar terms. With a sharp peak at the location of the first-order scalar and tensor power spectrum peaks, its effect on the overall induced SOGW spectrum is to smooth out the peaks in the Gaussian terms. This is a typical effect of non-Gaussian contributions. The HST term extends further into the UV than the Gaussian scalar-tensor contribution, in the same way that non-Gaussian scalar-scalar terms do. It also suffers from the same divergence issue as the scalar-tensor Gaussian term for wide primordial power spectra and is affected by parity violation in a similar way. These commonalities suggest that it may be infeasible for LISA to detect the presence of a non-Gaussian scalar-tensor sector in the SGWB.

One feature that does seem to distinguish the HST term from others already studied, however, is the presence of a pronounced `knee' extending over scales $3k_p\leq k \leq 4k_p$. This knee typically has a much larger amplitude than the knees generated by scalar-scalar non-Gaussian terms and, for suitably large values of $F_{\textrm{NL}}$ or a fortuitous position of the power spectrum peak, could lie within LISA's sensitivity. There is a possibility, therefore, that features in the UV tail of any detected induced SOGW signature could be used to indicate the existence of scalar PNG.

While we have only considered introducing scalar non-Gaussianity into the scalar-tensor sector, it would be intriguing to extend this approach to the study of the effects of primordial tensor non-Gaussianity on the spectral density of induced SOGWs. This would have an impact on the scalar-tensor term investigated in this work and also on the tensor-tensor sector. From a phenomenological perspective, however, it is more challenging —-- and arguably less general —-- to justify the local expansion of the tensor field in the same way we did for the scalar case. A more straightforward extension of our work would be to include higher order non-Gaussian parameters such as $G_{\textrm{NL}}$ in both the scalar-scalar and scalar-tensor sectors and to compare their effects.

\acknowledgments
The authors are thankful for useful discussions with David Mulryne and Karim Malik. RP is also thankful to Guillem Dom\`{e}nech for fruitful discussions and Gianmassimo Tasinato for comments on tensor non-Gaussianity. RP is funded by STFC grant ST/P000592/1 and MWD is supported by a studentship awarded by the Perren Bequest.
This research utilised Queen Mary's Apocrita HPC facility, supported by QMUL Research-IT~\cite{king_2017_438045}.

\appendix
\section{Polarisation tensors and functions}
\label{sec:Polarisation}
In spherical polar coordinates the wave-vector $\mathbf{k}$ along which a GW propagates is expressed as
\begin{equation}
\label{kSpherical}
    \mathbf{k}=k(\sin{\theta_k}\cos{\phi_k},\sin{\theta_k}\sin{\phi_k},\cos{\theta_k})\,
\end{equation}
where $k$ is the magnitude of the momentum and $\theta_k$ and $\phi_k$ are the polar and azimuthal angles respectively. With this choice of coordinates we can construct two orthonormal polarisation vectors that span the subspace perpendicular to $\mathbf{k}$
\begin{equation}
    \label{Polarisationvector1}
    \mathbf{e}(\mathbf{k})=(\cos{\theta_k}\cos{\phi_k},\cos{\theta_k}\sin{\phi_k},-\sin{\theta_k})\,,
\end{equation}
\begin{equation}
    \label{Polarisationvector2}
    \overline{\mathbf{e}}(\mathbf{k})=(-\sin{\phi_k},\cos{\phi_k},0)\,.
\end{equation}
These vectors satisfy the relations
\begin{eqnarray}
\label{PolProp}
    e_a(\mathbf{k})e^a(\mathbf{k})=\overline{e}_a(\mathbf{k}) \overline{e}^a(\mathbf{k})=1, \quad e_a(\mathbf{k})\overline{e}^a(\mathbf{k})=0, \quad e_a(\mathbf{k})k^a=\overline{e}_a(\mathbf{k})k^a=0 \,.
\end{eqnarray}
From these we can build polarisation tensors in the $\{+,\times\}$ basis
\begin{align}
\label{+xPolarisation}
    \begin{split}
         q_{ab}^{+}(\mathbf{k})&=\frac{1}{\sqrt{2}}\left ( e_a(\mathbf{k})e_b(\mathbf{k}) - \overline{e}_a(\mathbf{k}) \overline{e}_b(\mathbf{k})\right )\text{ ,} \\
        q_{ab}^{\times}(\mathbf{k})&=\frac{1}{\sqrt{2}}\left ( e_a(\mathbf{k})\overline{e}_b(\mathbf{k}) + \overline{e}_a(\mathbf{k}) e_b(\mathbf{k})\right )\,.
    \end{split}
\end{align}
In this paper we choose to work in terms of polarisation tensors in the circular $\{R,L\}$ basis instead. This simplifies the computation of products of projection factors defined in Eq.~\eqref{defpoldecomp}. The circular basis is constructed in terms of the polarisation tensors in Eq.~\eqref{+xPolarisation} as
\begin{align}
    \begin{split}
        q^R_{ab}(\mathbf{k}) &= \frac{1}{\sqrt{2}} \left ( q^+_{ab}(\mathbf{k}) + i q^{\times}_{ab}(\mathbf{k})  \right ) \text{ ,}  \\
        q^L_{ab}(\mathbf{k}) &= \frac{1}{\sqrt{2}} \left ( q^+_{ab}(\mathbf{k}) - i q^{\times}_{ab}(\mathbf{k})  \right )\,,
    \end{split}
\end{align}
which satisfy the traceless, transverse and normalisation conditions 
\begin{equation}
    q_{ab}^{\lambda}(\mathbf{k})\delta ^{ab} =0, \quad q_{ab}^{\lambda}(\mathbf{k}) k^a = 0, \quad \left( q_{ab}^{\lambda}(\mathbf{k}) \right)^* q^{ab,\lambda ^{\prime}}(\mathbf{k}) = \delta ^{\lambda \lambda ^{\prime}} \,,
\end{equation}
for $\lambda=R,L$. Furthermore we have the useful relations\footnote{This is explicitly clear in spherical coordinates, since $\mathbf{e}(\mathbf{k})$ and $\overline{\mathbf{e}}(\mathbf{k})$ are even and odd respectively under the transformation $\mathbf{k}\rightarrow -\mathbf{k}$. The same applies for $q_{ab}^{+}(\mathbf{k})$ and $q_{ab}^{\times}(\mathbf{k})$.}
\begin{equation}
    \left ( q_{ab}^{\lambda}(\mathbf{k}) \right)^* = q_{ab}^{-\lambda}(\mathbf{k})  =q_{ab}^{\lambda}(-\mathbf{k})\,,
\end{equation}
where $-\lambda$ refers to the opposite polarisation to $\lambda$, i.e. $-L=R$ and $-R=L$. 

Since most computations involve the interplay between two momenta, we also define the vector $\mathbf{p}$
\begin{equation}
    \mathbf{p}=p(\sin{\theta_p}\cos{\phi_p},\sin{\theta_p}\sin{\phi_p},\cos{\theta_p})\,.
\end{equation}
In Sec.~\ref{sec:changecoords} and App.~\ref{NGSIGWS} we transform coordinates from spherical polar to $v$ and $u$ defined in Eqs.~\eqref{ucoord} and~\eqref{vcoord}. Some useful relations include
\begin{equation}\label{reluv}
    \sin{\theta_p}=\sqrt{1-\left( \frac{1+v^2-u^2}{2v}\right)^2}\;,\quad \cos{\theta_p}=\frac{1+v^2-u^2}{2v}\, .
\end{equation}
In the scalar-scalar case, the product of polarisation functions is given by
\begin{equation}
     Q^{ss}_{\lambda}(\mathbf{k},\mathbf{p})Q^{ss}_{\lambda}(\mathbf{-k},\mathbf{-p})=\left (q^{ij}_{\lambda}(\mathbf{k}\right )^*p_ip_j\left ( q^{mn}_{\lambda}(\mathbf{-k}) \right )^* p_mp_n \, .
\end{equation}
Aligning $\mathbf{k}$ with the z-axis and using the relations in Eq.~(\ref{reluv}) we find that
\begin{equation}
     Q^{ss}_{\lambda}(\mathbf{k},\mathbf{p})Q^{ss}_{\lambda}(\mathbf{-k},\mathbf{-p})=\frac{1}{4}k^4v^4\left (1- \frac{\left(1-u^2+v^2 \right )^2}{4v^2} \right )^2\, ,
\end{equation}
for $\lambda=R,L$. In the scalar-tensor case, the polarisation functions are given by
\begin{align}
    \begin{split}
        &\sum _{\lambda _1} Q^{st}_{\lambda , \lambda _1}(\mathbf{k},\mathbf{p})Q^{st}_{\lambda , \lambda _1}(\mathbf{-k},-\mathbf{p})P^{\lambda_1}_h(p) \\
        =&\sum _{\lambda _1} \left( q^{ab}_{\lambda}(\mathbf{k}) \right)^*q^{\lambda _1}_{ab}(\mathbf{p}) \left( q^{mn}_{\lambda}(-\mathbf{k}) \right)^*q^{\lambda _1}_{mn}(-\mathbf{p})P^{\lambda_1}_h(p)\, \\
       &=
       \begin{cases}
        \frac{\left ( u^2 - (1+v)^2 \right )^4P_h^R(p)  + \left ( u^2 - (-1+v)^2 \right )^4P_h^L(p) }{256v^4}\, , \lambda = R \, \\
         \frac{\left ( u^2 - (1+v)^2 \right )^4P_h^L(p)  + \left ( u^2 - (-1+v)^2 \right )^4P_h^R(p) }{256v^4}\, ,\lambda =L 
    \end{cases}
    \,.
    \end{split}
\end{align}

\section{Scalar-Scalar non-Gaussian contributions}
\label{NGSIGWS}
Here we list the scalar-scalar contributions to the induced SOGW spectrum for a locally non-Gaussian scalar curvature perturbation~\eqref{localNGexpansion} up to $\mathcal{O}(F_{\textrm{NL}}^4)$. We begin by defining the scalar-scalar kernel in RD
\begin{equation}
    \mathcal{I}^{ss}(v,u,x)\equiv k^2 I^{ss}\left(kv,ku,1/\sqrt{3},x/k\right)\,,
\end{equation}
which is a slight modification of the original scalar-scalar kernel in Eq.~\eqref{kernelss}. In general, evaluating the contributions listed below will involve late-time limit, oscillation-averaged products of $\mathcal{I}^{ss}$~\cite{Adshead:2021hnm}.
\begin{equation}
\begin{split}
    \overline{\mathcal{I}^{ss}(v_1,u_1,x \rightarrow \infty)\mathcal{I}^{ss}(v_2,u_2,x \rightarrow \infty)}=\frac{1}{2x^2}&I_A(u_1,v_1)I_A(u_2,v_2)\\
    \times&\left[ I_B(u_1,v_1)I_B(u_2,v_2)+\pi^2 I_C(u_1,v_1)I_C(u_2,v_2)\right]\,,
\end{split}
\end{equation}
where
\begin{align}
    I_A(u,v)&=\frac{3(u^2+v^2-3)}{4u^3v^3}\,,\\
    I_B(u,v)&=-4uv+(u^2+v^2-3)\log\Bigg|\frac{3-(u+v)^2}{3-(u-v)^2}\Bigg|\,, \\
    I_C(u,v)&=(u^2+v^2-3)\,\Theta(v+u-\sqrt{3})\,.
\end{align}
Further defining
\begin{equation}
    \mathcal{J}_{ss}(v,u,x)\equiv v^2 \sin^2 \theta \,\mathcal{I}^{ss}(v,u,x)\,,
\end{equation}
where $\theta$ is the polar angle of the momentum $\mathbf{p}$ in $Q_\lambda^{ss}(\mathbf{k},\mathbf{p})$ (see Eq.~\eqref{reluv}), we can now list the polarisation-summed, oscillation-averaged dimensionless power spectra that contribute to the induced SOGW spectrum in compact forms. These include the scalar-scalar Gaussian term
\begin{equation}
    \overline{\mathcal{P}^{ss}(\eta,k)_{\textrm{Gaussian}}}=\int_0^\infty \D t \, \int^1_{-1} \D s \, \frac{1}{u^2 v^2} \overline{\mathcal{J}_{ss}(k,v,u,c_s)^2} \mathcal{P}_{\mathcal{R},\textrm{G}}(ku) \mathcal{P}_{\mathcal{R},\textrm{G}}(kv)\,,
\end{equation}
a disconnected hybrid term of $\mathcal{O}(F_{\textrm{NL}}^2)$
\begin{equation}
\begin{split}
    \overline{\mathcal{P}^{ss}(\eta,k)_{\textrm{hybrid}}}=F_{\textrm{NL}}^2\prod_{i=1}^{2}&\left[\int_0^\infty \D t_i \, \int^1_{-1} \D s_i \, \frac{1}{u_i^2 v_i^2}\right] \overline{\mathcal{J}_{ss}(k,v_1,u_1,c_s)^2} \\
    &\times \mathcal{P}_{\mathcal{R},\textrm{G}}(ku_1) \mathcal{P}_{\mathcal{R},\textrm{G}}(kv_1 v_2)\mathcal{P}_{\mathcal{R},\textrm{G}}(ku_2 v_1)\,,
\end{split}
\end{equation}
two connected $\mathcal{O}(F_{\textrm{NL}}^2)$ terms
\begin{equation}
\begin{split}
    \overline{\mathcal{P}^{ss}(\eta,k)_{\textrm{C}}}=\frac{F_{\textrm{NL}}^2}{\pi}\prod_{i=1}^{2}&\left[\int_0^\infty \D t_i \, \int^1_{-1} \D s_i \,\overline{\mathcal{I}_{ss}(k,v_i,u_i,c_s)}\right] \, \int^{2\pi}_{0} \D \varphi \, \cos{2\varphi} \\
    &\times\frac{\mathcal{P}_{\mathcal{R},\textrm{G}}(kv_2)}{v_2^3} \frac{\mathcal{P}_{\mathcal{R},\textrm{G}}(k u_2)}{u_2^3}\frac{\mathcal{P}_{\mathcal{R},\textrm{G}}(k w_a)}{w_a^3}\,,
\end{split}
\end{equation}
\begin{equation}
\begin{split}
    \overline{\mathcal{P}^{ss}(\eta,k)_{\textrm{Z}}}=\frac{F_{\textrm{NL}}^2}{\pi}\prod_{i=1}^{2}&\left[\int_0^\infty \D t_i \, \int^1_{-1} \D s_i \,\overline{\mathcal{I}_{ss}(k,v_i,u_i,c_s)}\right]\, \int^{2\pi}_{0} \, \D \varphi \cos{2\varphi}\,\\
    &\times\frac{\mathcal{P}_{\mathcal{R},\textrm{G}}(kv_1)}{v_1^3} \frac{\mathcal{P}_{\mathcal{R},\textrm{G}}(k v_2)}{v_2^3}\frac{\mathcal{P}_{\mathcal{R},\textrm{G}}(k w_b)}{w_b^3}\,,
\end{split}
\end{equation}
where
\begin{equation}
    w_a^2=v_1^2+v_2^2-2\frac{\mathbf{q}_1 \cdot \mathbf{q}_2}{k^2}\,,
\end{equation}
\begin{equation}
    w_b^2=1+v_1^2+v_2^2-2\frac{\mathbf{k} \cdot \mathbf{q}_1}{k^2}-2\frac{\mathbf{k} \cdot \mathbf{q}_2}{k^2}+2\frac{\mathbf{q}_1 \cdot \mathbf{q}_2}{k^2}\,,
\end{equation}
and the dot-products between momenta are expressed in terms of the integration variables as
\begin{equation}
    \begin{split}
        \frac{\mathbf{q}_i \cdot \mathbf{q}_j}{k^2}=&\frac{\cos \phi}{4}\sqrt{t_i(t_i+2)(1-s_i^2)t_j(t_j+2)(1-s_j^2)}\\
        &+\frac{1}{4}\left[1-s_i(t_i+1)\right]\left[1-s_j(t_j+1)\right]\,,
    \end{split}
\end{equation}
and
\begin{equation}
    \frac{\mathbf{k} \cdot \mathbf{q}_i}{k^2}= \frac{1}{2}\left[1-s_i(t_i+1)\right]\,.
\end{equation}
At $\mathcal{O}(F_{\textrm{NL}}^4)$ we have the disconnected term
\begin{equation}
\begin{split}
    \overline{\mathcal{P}^{ss}(\eta,k)_{\textrm{reducible}}}=\frac{F_{\textrm{NL}}^4}{4}\prod_{i=1}^{3}&\left[\int_0^\infty \D t_i \, \int^1_{-1} \D s_i \, \frac{1}{u_i^2 v_i^2}\right] \, \overline{\mathcal{I}_{ss}(k,v_1,u_1,c_s)^2}\\
    &\times\mathcal{P}_{\mathcal{R},\textrm{G}}(kv_1v_2) \mathcal{P}_{\mathcal{R},\textrm{G}}(k u_2 v_1) \mathcal{P}_{\mathcal{R},\textrm{G}}(k u_1 v_3) \mathcal{P}_{\mathcal{R},\textrm{G}}(k u_1 u_3)\,,
\end{split}
\end{equation}
and two connected terms
\begin{equation}
\begin{split}
    \overline{\mathcal{P}^{ss}(\eta,k)_{\textrm{planar}}}=\frac{F_{\textrm{NL}}^4}{4\pi^2}\prod_{i=1}^{3}&\left[\int_0^\infty \D t_i \, \int^1_{-1} \D s_i \, u_i v_i\right] \prod_{i=1}^{2}\left[\overline{\mathcal{I}_{ss}(k,v_i,u_i,c_s)}\right]\, \int^{2\pi}_{0} \D \varphi_{12}\, \int^{2\pi}_{0}  \D \varphi_{23}\\
    &\times\cos{2\varphi_{12}}\,\frac{\mathcal{P}_{\mathcal{R},\textrm{G}}(kv_3)}{v_3^3} \frac{\mathcal{P}_{\mathcal{R},\textrm{G}}(k w_{13})}{w_{13}^3}\frac{\mathcal{P}_{\mathcal{R},\textrm{G}}(k w_{23})}{w_{23}^3}\frac{\mathcal{P}_{\mathcal{R},\textrm{G}}(k u_3)}{u_3^3}\,,
\end{split}
\end{equation}
\begin{equation}
\begin{split}
    \overline{\mathcal{P}^{ss}(\eta,k)_{\textrm{non-planar}}}=\frac{F_{\textrm{NL}}^4}{8\pi^2}\prod_{i=1}^{3}&\left[\int_0^\infty \D t_i \, \int^1_{-1} \D s_i \, u_i v_i\right] \prod_{i=1}^{2}\left[\overline{\mathcal{I}_{ss}(k,v_i,u_i,c_s)}\right]\, \int^{2\pi}_{0} \D \varphi_{12}\, \int^{2\pi}_{0}  \D \varphi_{23}\\
    &\times\cos{2\varphi_{12}}\,\frac{\mathcal{P}_{\mathcal{R},\textrm{G}}(kv_3)}{v_3^3} \frac{\mathcal{P}_{\mathcal{R},\textrm{G}}(k w_{13})}{w_{13}^3}\frac{\mathcal{P}_{\mathcal{R},\textrm{G}}(k w_{23})}{w_{23}^3}\frac{\mathcal{P}_{\mathcal{R},\textrm{G}}(k w_{123})}{w_{123}^3}\,,
\end{split}
\end{equation}
where
\begin{equation}
    w_{i3}^2=v_i^2+v_3^2-2\frac{\mathbf{q}_i \cdot \mathbf{q}_3}{k^2}\,,
\end{equation}
\begin{equation}
\begin{split}
    w_{123}^2=1&+v_1^2+v_2^2+v_3^2-2\frac{\mathbf{k} \cdot \mathbf{q}_1}{k^2}-2\frac{\mathbf{k} \cdot \mathbf{q}_2}{k^2}+2\frac{\mathbf{k} \cdot \mathbf{q}_3}{k^2}\\
    &-2\frac{\mathbf{q}_1 \cdot \mathbf{q}_3}{k^2}-2\frac{\mathbf{q}_2 \cdot \mathbf{q}_3}{k^2}+2\frac{\mathbf{q}_1 \cdot \mathbf{q}_2}{k^2}\,.
\end{split}
\end{equation}
The integrals within each of these terms are computed using the numerical integration package \texttt{vegas+}~\cite{Lepage:2020tgj}. The terms are evaluated for 700 different external momenta, $k$, using Queen Mary's Apocrita HPC facility~\cite{king_2017_438045}. 
\section{Diagrammatic rules for scalar-scalar and scalar-tensor contributions}
\label{app:feynman}
In Ref.~\cite{Adshead:2021hnm}, Adshead et. al. present a comprehensive set of diagrammatic rules for the non-Gaussian scalar-scalar contributions to the induced SOGW spectrum. This is an alternative approach to the direct computation of the various terms arising from the insertion of a locally non-Gaussian scalar curvature perturbation~\eqref{localNGexpansion} which can be tedious. It has already been extended in Refs.~\cite{Perna:2024ehx,Li:2023xtl} to include higher-order local non-Gaussian expansion parameters such as $G_{\textrm{NL}}$ and $H_{\textrm{NL}}$. In this appendix, we reproduce the relevant scalar-scalar diagrammatic rules from Ref.~\cite{Adshead:2021hnm} for the construction of diagrams corresponding to the terms listed in App.~\ref{NGSIGWS} alongside introducing new scalar-tensor rules:
\begin{enumerate}
    \item[(i)] \quad \raisebox{-.45\height}{\includegraphics[width=.25\textwidth]{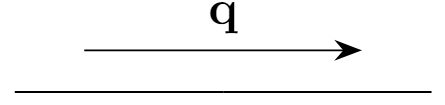}}$=\qquad P_{\mathcal{R},G}(q)$
    \item[(ii)] \quad \raisebox{-.45\height}{\includegraphics[width=.25\textwidth]{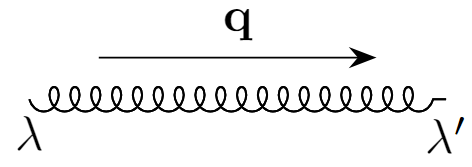}}$=\qquad \delta^{\lambda C} P_{h,C}(q)\delta^{C \lambda'}$
    \item[(iii)] \quad \raisebox{-.45\height}{\includegraphics[width=.25\textwidth]{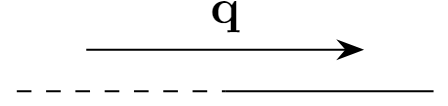}}$=\qquad 1$
    \item[(iv)] \quad \raisebox{-.45\height}{\includegraphics[width=.25\textwidth]{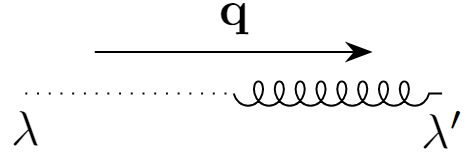}}$=\qquad \delta^{\lambda \lambda'}$
    \item[(v)] \quad \raisebox{-.45\height}{\includegraphics[width=.25\textwidth]{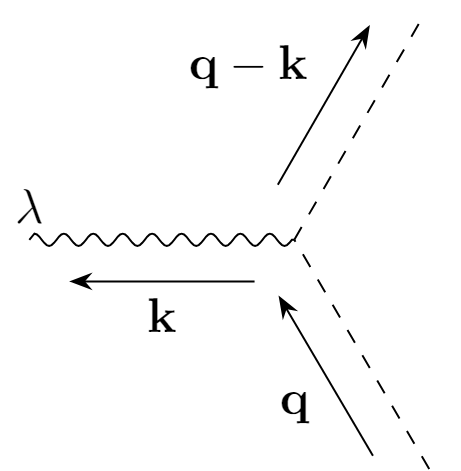}} $=\quad\displaystyle 2 \int_{0}^{\eta} d \bar{\eta} \,
        \frac{a(\bar{\eta})}{a(\eta)}
        G_\mathbf{k}(\eta, \bar{\eta})
        Q^{ss}_\lambda(\mathbf{k}, \mathbf{q})
        f^{ss}(c_s\eta p, c_s\eta |\ki - \qi|)$
    \item[(vi)] \raisebox{-.45\height}
    {\includegraphics[width=.3\textwidth]{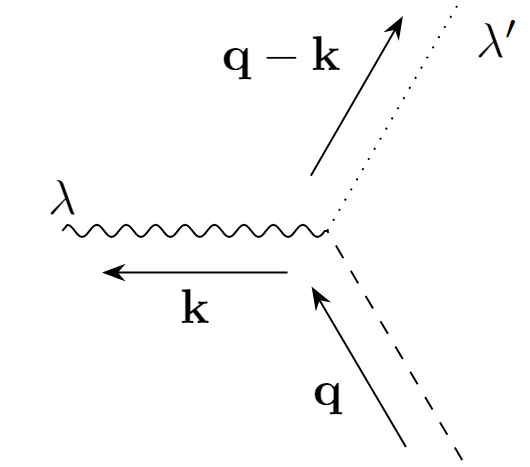}} $=\quad\displaystyle 4 \int_{0}^{\eta} d \bar{\eta} \,
        \frac{a(\bar{\eta})}{a(\eta)}
        G_\mathbf{k}(\eta, \bar{\eta})
        Q^{st}_{\lambda \lambda'}(\mathbf{k}, \mathbf{q})
        f^{st}(c_s\eta|\ki-\qi|, \eta q)$
     \item[(vii)] \quad \raisebox{-.45\height}{\includegraphics[width=.24\textwidth]{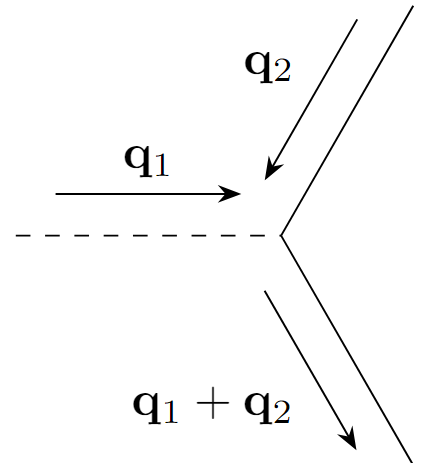}} $=\qquad F_{\textrm{NL}}$   
\end{enumerate}
The diagrams shown in Fig.~\ref{feynmandiagrams} were generated from these rules and can be used to derive the expressions~\eqref{stG} and~\eqref{stHS}. Wavy lines denote external SOGWs, solid lines are scalar power spectra, spiral lines are tensor power spectra, dashed lines represent the scalar-scalar transfer function and dotted lines represent the scalar-tensor transfer function. The letter $C$ is a specific GW polarisation, either $R$ or $L$. This means that each scalar-tensor diagram really represents two distinct contributions --- one for each value of $C$. Note that our rule for the scalar-scalar transfer function differs from that of Ref.~\cite{Adshead:2021hnm} due to our definition of the second-order tensor perturbation in Eq.~\eqref{PerturbedFLRW}.

\bibliography{refs} 
\bibliographystyle{JHEP}

\end{document}